\documentclass[aps,prd,twoside,twocolumn,superscriptaddress,floatfix,nofootinbib,showpacs]{revtex4}
\usepackage{amsmath}
\usepackage{bm}
\usepackage{xcolor}

\newcommand\beq{\begin{equation}}
\newcommand\eeq{\end{equation}}
\newcommand\beqn{\begin{eqnarray}}
\newcommand\eeqn{\end{eqnarray}}

\newcommand\bl{{\bm{\ell}}}

\newcommand\nn{\nonumber}

\def\x{{\bf x}}

\def\v{{\bf v}}
\def\y{{\bf y}}

\def\p{{\bf p}}

\def\l{{\bf  \ell}}
\def\L{{\bf L}}
\def\T{{\Theta}}
\def\Tt{\tilde{\Theta}}
\def\Tf{\tilde{\Theta}_f}

\def\l{{\bm \ell}}

\newcommand{\ba}{\begin{eqnarray}}
\newcommand{\ea}{\end{eqnarray}}
\newcommand{\be}{\begin{equation}}
\newcommand{\ee}{\end{equation}}

\newcommand\lsim{\mathrel{\rlap{\lower4pt\hbox{\hskip1pt$\sim$}}
        \raise1pt\hbox{$<$}}}
\newcommand\gsim{\mathrel{\rlap{\lower4pt\hbox{\hskip1pt$\sim$}}
        \raise1pt\hbox{$>$}}}

\newcommand{\jcap}{{J.~Cosm.~Astrop.~Phys.}}
\newcommand{\araa}{{Annu.~Rev.~Astron.~Astrophys.}}
\newcommand{\aap}{{Astron.~Astrophys.}}
\newcommand{\apjl}{{Astrophys.~J.~Lett.}}
\newcommand{\apjs}{{Astrophys.~J.~Supp.}}
\newcommand{\aj}{{Astron.~J.}}
\newcommand{\mnras}{{Mon.~Not.~R.~Astron.~Soc.}}
\usepackage{graphicx}
\usepackage[large]{subfigure}
\usepackage{amssymb, amsmath}
\usepackage[amssymb]{SIunits}
\usepackage{epstopdf}
\usepackage{aas_macros}
\usepackage{natbib}

\begin{document}

\title{The Kinematic Sunyaev-Zel'dovich Effect with Projected Fields II: prospects, challenges, and comparison with simulations}
 \author{Simone~Ferraro}\email{sferraro@berkeley.edu}
 \affiliation{Berkeley Center for Cosmological Physics and
Department of Astronomy, University of California, Berkeley, CA, USA 94720}
  \affiliation{Miller Institute for Basic Research in Science, University of California, Berkeley, CA, 94720, USA}
  \author{J.~Colin~Hill}
 \affiliation{Dept.~of Astronomy, Pupin Hall, Columbia University, New York, NY USA 10027}
  \author{Nick~Battaglia}
 \affiliation{Dept.~of Astrophysical Sciences, Peyton Hall, Princeton University, Princeton, NJ USA 08544}
 \author{Jia~Liu}
 \affiliation{Dept.~of Astronomy, Pupin Hall, Columbia University, New York, NY USA 10027} 
 \author{David~N.~Spergel}
 \affiliation{Dept.~of Astrophysical Sciences, Peyton Hall, Princeton University, Princeton, NJ USA 08544}
\begin{abstract}

The kinematic Sunyaev-Zel'dovich (kSZ) signal is a powerful probe of the cosmic baryon distribution.  The kSZ signal is proportional to the integrated free electron momentum rather than the electron pressure (which sources the thermal SZ signal).  Since velocities should be unbiased on large scales, the kSZ signal is an unbiased tracer of the large-scale electron distribution, and thus can be used to detect the ``missing baryons" that evade most observational techniques.

While most current methods for kSZ extraction rely on the availability of very accurate redshifts, we revisit a method that allows measurements even in the absence of redshift information for individual objects. It involves cross-correlating the \emph{square} of an appropriately filtered cosmic microwave background (CMB) temperature map with a projected density map constructed from a sample of large-scale structure tracers.
We show that this method will achieve high signal-to-noise when applied to the next generation of high-resolution CMB experiments, provided that component separation is sufficiently effective at removing foreground contamination.
Considering statistical errors only, we forecast that this estimator can yield $S/N \approx $ 3, 120 and over 150 for \emph{Planck}, \emph{Advanced ACTPol}, and a hypothetical Stage-IV CMB experiment, respectively, in combination with a galaxy catalog from \emph{WISE}, and about 20\% larger $S/N$ for a galaxy catalog from the proposed \emph{SPHEREx} experiment.  We show that the basic estimator receives a contribution due to leakage from CMB lensing, but that this term can be effectively removed by either direct measurement or marginalization, with little effect on the kSZ significance. We discuss possible sources of systematic contamination and propose mitigation strategies for future surveys. We compare the theoretical predictions to numerical simulations and validate the approximations in our analytic approach.

This work serves as a companion paper to the first kSZ measurement with this method, where we used CMB temperature maps constructed from \emph{Planck} and \emph{WMAP} data, together with galaxies from the \emph{WISE} survey, to obtain a 3.8 - 4.5$\sigma$ detection of the kSZ$^2$ amplitude. 

\end{abstract}
\maketitle

\section{Introduction}
The amount of baryonic matter in the Universe is tightly constrained at high redshift by measurements of the primordial cosmic microwave background (CMB) anisotropies~\cite{Hinshawetal2013,2015arXiv150201582P} and of the abundance of light elements formed through the process of Big Bang nucleosynthesis (BBN) \cite{Steigman2007}.  The baryonic abundance of the present-day Universe must satisfy these primordial constraints, assuming the absence of unknown, exotic physics.  However, the cosmic baryon census at low redshifts has long fallen short of the expected value~(e.g.,~\cite{Fukugitaetal1998,Bregman2007}), especially for halos smaller than galaxy clusters, such as individual galaxies or groups of galaxies.  One hypothesis is that these ``missing baryons'' reside in an ionized, diffuse component known as the Warm-Hot Intergalactic Medium \cite{2006ApJ...650..560C}, which has been difficult to detect in X-ray emission due to its relatively low density and temperature.
Observations of highly ionized gas in quasar absorption lines provide some evidence and constraints on its properties \cite{2014ApJ...792....8W, 2016MNRAS.tmp...78B}.

The kinematic Sunyaev-Zel'dovich (kSZ) effect is caused by Compton-scattering of CMB photons off of free electrons moving with a non-zero line-of-sight (LOS) velocity~\cite{Sunyaev-Zeldovich1972,Sunyaev-Zeldovich1980,Ostriker-Vishniac1986}.  The corresponding shift in the observed CMB temperature is proportional to both the total number of electrons (or optical depth) and their LOS velocity, which is equally likely to be positive or negative.  Moreover, the kSZ signal should be unbiased, in the sense that halos of different masses move in the same large-scale cosmic velocity field, and therefore it is a direct probe of the electron density.  Thus it can be used to measure the ionized gas abundance and distribution in galaxies and clusters. These measurements can be performed as a function of mass and redshift (and other galaxy properties of interest), informing us about the extent and nature of feedback processes.

If the cluster optical depth can be determined through other methods, the kSZ effect can be used to measure statistics of LOS velocities, which are sensitive to the rate of growth of structure and are hence a powerful probe of dark energy or modified gravity \cite{2008PhRvD..77h3004B}.

The kSZ effect was first detected in \emph{Atacama Cosmology Telescope} (\emph{ACT}) data by studying the pairwise momenta of luminous galaxies in the Baryon Oscillation Spectroscopic Survey (BOSS) DR9 catalog~\cite{Handetal2012}. Recent analyses of the \emph{Planck}, \emph{ACTPol} and \emph{South Pole Telescope} (\emph{SPT-SZ}) datasets have found additional evidence for the signal, using large-scale structure catalogs from the Sloan Digital Sky Survey (SDSS) and Dark Energy Survey (DES) \cite{2015arXiv150403339P, 2015PhRvL.115s1301H, 2015arXiv151006442S, 2016arXiv160303904S}. A high-resolution analysis of a particular galaxy cluster also found evidence for the kSZ effect in that system~\cite{Mroczkowskietal2012,Sayersetal2013}.  

Most kSZ estimators in the literature \cite{2009arXiv0903.2845H, 2011MNRAS.413..628S, 2014MNRAS.443.2311L, 1999ApJ...515L...1F, 2008PhRvD..77h3004B} require spectroscopic redshifts.  The use of photometric redshifts leads to a large degradation in the statistical significance of the kSZ detection \cite{2013ApJ...765L..32K, 2015arXiv151102843F}.  In this paper, we revisit a method that only makes use of projected fields and therefore does not require individual redshifts for each object, but only a statistical redshift distribution for the low-redshift tracers used in the analysis.  Such a distribution could be constructed from photometric redshift data, but even photometric redshifts are not necessarily required --- a well-understood sub-sample cross-matched to existing redshift catalogs would suffice. The main motivation of this estimator is that photometric or imaging surveys are much cheaper than their spectroscopic counterparts and are able to map larger volumes of the Universe. An excellent example is the \emph{Wide-field Infrared Survey Explorer} (\emph{WISE}) data set~\cite{2010AJ....140.1868W}, which covers the full sky in the mid-infrared. Moreover, the kSZ technique described here will have comparable statistical power and yield independent information to the traditional methods when applied to future high-resolution CMB experiments, if component separation allows an effective removal of frequency-dependent foregrounds.

The basic idea behind this estimator is that because of the equal likelihood of positive and negative kSZ signals, an appropriately filtered version of the CMB temperature map must be squared in real space before cross-correlating with tracers (e.g., galaxies, quasars, or gravitational lensing convergence); we thus refer to this as the kSZ${}^2$--tracer cross-correlation.  Crucially, the CMB temperature map must be cleaned of foreground (non-kSZ) emission associated with the tracer objects, and thus a multi-frequency analysis is necessary.  First suggested in~\cite{Doreetal2004} and studied further in~\cite{DeDeoetal2005}, the kSZ${}^2$--tracer cross-correlation probes the mass and LOS velocity of the ionized gas associated with the tracer objects in the large-scale structure sample.  In other words, the CMB temperature itself contains kSZ information, and this is just the lowest-order non-zero estimator that allows one to extract the signal from a given tracer population without requiring 3D information. This is in essence a measurement of the \emph{squeezed} limit of the bispectrum of two powers of the CMB temperature and one power of the projected tracer field, and it can be shown to be the configuration containing most of the information (we leave a full treatment of optimality to future work).

Because the estimator is quadratic in temperature, it is affected by leakage from weak lensing of the CMB, and this lensing contribution --- which can be larger than the signal in some instances --- must be appropriately removed or marginalized over. Fortunately, the multipole-dependence of the lensing leakage is quite different than the kSZ${}^2$ signal, and thus it can be marginalized with very little effect on the statistical significance of the kSZ${}^2$ signal.

We have recently presented the first measurement of the baryon abundance with this technique in a companion paper \cite{2016arXiv160301608H} (hereafter H16). We used a galaxy catalog constructed from \emph{WISE} data~\cite{2010AJ....140.1868W} and CMB temperature maps cleaned via ``local-generalized morphological component analysis'' (LGMCA)~\cite{Bobinetal2015} constructed from the \emph{Planck} full mission~\cite{2015arXiv150201582P} and \emph{Wilkinson Microwave Anisotropy Probe} (\emph{WMAP}) nine-year survey (WMAP9) data~\cite{Bennettetal2013}.  We detected the kSZ$^2$ signal with signal-to-noise ($S/N$) $\approx 3.8 - 4.5$, depending on the use of external CMB lensing information, and thus obtained a 13\% measurement of the baryon abundance at $z \approx 0.4$.

Except where explicitly stated otherwise, we use cosmological parameters from the 2015 \emph{Planck} data release~\cite{Planck2015params}.

The remainder of this paper is organized as follows: in Section \ref{sec:theory} we review the theory, including the approximations in our analytic approach. In Section \ref{sec:forecast} we present forecasts for current and future experiments, while Section \ref{sec:lensing} discusses the lensing contribution and ways to remove it.  In Section \ref{sec:sims}, we present a comparison of the theory with numerical simulations to check the accuracy of our approximations. We discuss our recent measurement using this method in Section \ref{sec:H16}.
Foreground contamination poses a serious challenge for this type of measurement, which we discuss in Section \ref{sec:challenges}. We conclude in Section \ref{sec:conclusions}.

\section{Theory}
\label{sec:theory}
The kSZ effect produces a CMB temperature change, $\Theta^{\rm kSZ}(\hat{\mathbf{n}}) = \Delta T^{\rm kSZ}/T_{\rm CMB}(\hat{\mathbf{n}})$, in a direction $\hat{\mathbf{n}}$ on the sky (in units with $c$ = 1):
\ba
\label{eq.kSZdef}
\Theta^{\rm kSZ}(\hat{\mathbf{n}}) & = & - \int d\eta \ g(\eta) \ \mathbf{p}_e \cdot \mathbf{\hat{n}} \\
& = & - \sigma_T \int \frac{d \eta}{1+z} e^{-\tau} n_e(\hat{\mathbf{n}},\eta) \ \mathbf{v}_e \cdot \mathbf{\hat{n}} \,,
\label{eq.kSZdef2}
\ea
where $\sigma_T$ is the Thomson scattering cross-section, $\eta(z)$ is the comoving distance to redshift $z$, $\tau$ is the optical depth to Thomson scattering, $g(\eta) = \dot{\tau} e^{-\tau}$ is the visibility function, $n_e$ is the physical free electron number density, $\mathbf{v}_e$ is the peculiar velocity of the electrons, and we have defined the electron momentum $\mathbf{p}_e = (1+\delta_e) \mathbf{v}_e$.

For concreteness, we consider galaxies as tracers in the following, but the formalism extends straightforwardly to any other tracer of the late-time density field (such as quasars, lensing convergence, or 21 cm fluctuations).

The projected galaxy overdensity $\delta_g$ is given by
\be
\delta_g (\hat{\mathbf{n}}) =  \int_0^{\eta_{\rm max}} d \eta \ W^{g}(\eta) \ \delta_m(\eta \hat{\mathbf{n}}, \eta) \,,
\label{eq:delta_gproj}
\ee
where $\eta_{\rm max}$ is the maximum source distance, $\delta_m$ is the (three-dimensional) matter overdensity, and $W^g(\eta)$ is the projection kernel:
\be
W^{g} (\eta) = b_g p_s(\eta) \,.
\label{eq.Wg}
\ee
Here $p_s(\eta) \propto dn/d\eta$ is the redshift distribution of the galaxies (normalized to have unit integral) and $b_g$ is the linear galaxy bias.

As explained in the introduction, the cross-correlation between the kSZ signal and low-redshift tracers is expected to vanish on small scales (where the contribution from the integrated Sachs-Wolfe (ISW) effect is expected to be negligible) because of the $\mathbf{v}_e \rightarrow - \mathbf{v}_e$ symmetry. We therefore square the CMB temperature fluctuation map in real space before cross-correlating it with a tracer density map.

In order to downweight angular scales dominated by noise (in our case primary CMB fluctuations and detector noise), we filter the temperature map in harmonic space with a Wiener filter $F$ before squaring in real space:
\be
\label{eq.Felldef}
F(\ell) = \frac{C_{\ell}^{\rm kSZ}}{C_{\ell}^{\rm tot}} \,,
\ee
where $C_{\ell}^{\rm kSZ}$ is the (theoretical) kSZ power spectrum and $C_{\ell}^{\rm tot}$ is the total fluctuation power, which includes primary CMB, kSZ, ISW, noise, and any residual foregrounds.  Our template for $C_{\ell}^{\rm kSZ}$ in Equation~\ref{eq.Felldef} is derived from cosmological hydrodynamics simulations~\cite{BBPSS2010}.

Moreover, the CMB is observed through a finite beam $b(\ell)$, so that the total filtered map $\Theta_f$ is related to the underlying (true) CMB anisotropy $\Theta$ by
\be
\Theta_f(\boldsymbol{\ell}) = F(\ell) b(\ell) \Theta(\boldsymbol{\ell}) \equiv f(\ell) \Theta(\boldsymbol{\ell})
\ee
where we have defined $f(\ell) = F(\ell) b(\ell)$.

In this work, we are interested in the cross-correlation $C_\ell^{\rm{kSZ}^2 \times \delta_g}$ between the square of the filtered CMB map and tracers:
\be
\langle \Theta_f^2(\l) \delta_g(\l') \rangle  = (2 \pi)^2 \delta_D(\l + \l') \ C_\ell^{\rm{kSZ}^2 \times \delta_g} \,.
\ee
Following \cite{Doreetal2004, DeDeoetal2005} we can write the angular power spectrum of the kSZ${}^2$--galaxy cross-correlation as
\be
C_\ell^{\rm{kSZ}^2 \times \delta_g} = \int_0^{\eta_{\rm max}} \frac{d \eta} {\eta^2} W^{g}(\eta) g^2(\eta) \mathcal{T}(k = \ell / \eta, \eta) \,,
\label{eq:Cltheory}
\ee
where we have used the Limber approximation \cite{1953ApJ...117..134L}, and the \textit{triangle power spectrum} $\mathcal{T}$
\be
\mathcal{T}(k, \eta) = \int \frac{d^2 \mathbf{q}}{(2 \pi)^2} f(q \eta) f(|\mathbf{k} + \mathbf{q}|\eta) B_{\delta p_{\hat{\mathbf{n}}} p_{\hat{\mathbf{n}}} }(\mathbf{k}, \mathbf{q}, -\mathbf{k} -\mathbf{q}) \,.
\ee
Here, the hybrid bispectrum $B_{\delta p_{\hat{\mathbf{n}}} p_{\hat{\mathbf{n}}} }$ is the three-point function of one density contrast and two LOS electron momenta, $p_{\hat{\mathbf{n}}}$.  The triangle power spectrum $\mathcal{T}$ is the integral over all triangles with sides $\mathbf{k}$, $\mathbf{q}$, and $ -\mathbf{k} -\mathbf{q}$, lying on planes of constant redshift.  Since the momentum field is $\p \sim \v \delta$ on small scales, the hybrid bispectrum $B_{\delta p_{\hat{\mathbf{n}}} p_{\hat{\mathbf{n}}} }$ is the sum of terms of the form $\langle vv \rangle \langle \delta \delta \delta \rangle$, $\langle v \delta \rangle \langle \delta \delta v \rangle$, etc., and a connected part $\langle vv \delta \delta \delta \rangle_c$. Ref.~\cite{DeDeoetal2005} argues that the former term $\langle vv \rangle \langle \delta \delta \delta \rangle$ dominates on small scales ($k \gg k_{\rm eq}$) and we will assume that the non-Gaussianity is weak enough that the connected part can be neglected.

On small scales we can therefore approximate the hybrid bispectrum in terms of the 3D velocity dispersion $v^2_{\rm rms}$ and the non-linear matter bispectrum $B_m^{\rm NL}$  \cite{Doreetal2004, DeDeoetal2005}:
\be
B_{\delta p_{\hat{\mathbf{n}}} p_{\hat{\mathbf{n}}} }\approx \frac{1}{3} v^2_{\rm rms} B_m^{\rm NL}
\label{bispectrum}
\ee
We use fitting functions from \cite{GilMarin:2011ik} for the non-linear matter bispectrum $B_m^{\rm NL}$ and the velocity dispersion $v^2_{\rm rms}$ is computed in linear theory, which should be an excellent approximation.\footnote{Numerical simulations~\cite{Hahn:2014lca} show that linear theory is a very good approximation to the velocity power spectrum up to $k \approx 0.5 \, h$/Mpc, and therefore the velocity dispersion, which receives most of its contribution from larger scales, should be well approximated by linear theory.}  We test the validity of the approximations made here by comparison to numerical simulations in Section~\ref{sec:sims}, and we find that these are excellent on the scales relevant for the analysis of a \emph{Planck}-like experiment.

At late times, some fraction of the cosmological abundance of electrons lies in stars or neutral media and therefore does not take part in the Thomson scattering that produces the kSZ signal. We define $f_{\rm free}$ as the fraction of free electrons, and note that in general this quantity will be redshift-dependent. The visibility function $g(\eta)$ in Equation~\ref{eq.kSZdef} is proportional to $f_{\rm free}$, so that $C_\ell^{\rm{kSZ}^2 \times \delta_g}$ scales like $f_{\rm free}^2$ and hence can be used to measure the free electron fraction.  In H16 we note that the signal is also proportional to the (square of the) baryon fraction $f_b = \rho_b /\rho_m$, so that if we allow $f_b$ to vary, the amplitude of $C_\ell^{\rm{kSZ}^2 \times \delta_g}$ provides a measurement of the product $f_{\rm free} f_b$. For convenience in what follows we will fix $f_b = 0.155$, the fiducial value in our assumed cosmology.

Technically, the bispectrum in Equation \ref{bispectrum} is the three-point function of one matter and two electron overdensities, but for the purpose of forecasts, we will assume that the free electrons trace the dark matter down to the scales of interest.
While this is expected to be true for an experiment with the resolution of \emph{Planck}, this assumption will not hold as experiments proceed to higher resolution. The overall amplitude of the signal is set by $f_{\rm free}$, but the shape of the cross-correlation on small scales is directly related to the baryon profiles around galaxies and clusters, which are expected to be heavily influenced by feedback processes (for a measurement of the kSZ signal as a function of scale for group-size tracers see \cite{2015arXiv151006442S}).

\section{Forecasts}
\label{sec:forecast}
In this section we present forecasts for detection of the kSZ$^2$ signal. As discussed above, the amplitude of $C_\ell^{\rm{kSZ}^2 \times \delta_g}$ is proportional to the galaxy bias $b_g$ so that we can define
\be
\left( C_\ell^{\rm{kSZ}^2 \times \delta_g} \right)_{\rm measured} = b_g \mathcal{A}_{\rm kSZ^2} \ \left( C_\ell^{\rm{kSZ}^2 \times \delta_g} \right)_{\rm fiducial}
\ee
where the fiducial prediction assumes unit galaxy bias and full ionization, such that $\mathcal{A}_{\rm kSZ^2} \propto f_{\rm free}^2$.
It is often the case that the galaxy bias is either known externally to high accuracy (for example from the auto-correlation function or in cross-correlation with CMB lensing maps), or absent (for example if our tracer were lensing convergence). Therefore, in this section we will assume that we have an external sharp prior on the bias, so that the fractional error on $(b_g \mathcal{A}_{\rm kSZ^2}$) is the same as on $\mathcal{A}_{\rm kSZ^2}$. If this is not the case, we will show that the bias can be jointly fit together with $\mathcal{A}_{\rm kSZ^2}$, thanks to the fact that there is a lensing contribution to the measured $C_\ell^{\rm{kSZ}^2 \times \delta_g}$ which is proportional to $b_g$, but independent of the kSZ amplitude, as explained in Section~\ref{sec:lensing}. If the galaxy bias is obtained by a joint fit, there will be some (generally small) degradation in significance that depends on the experimental configuration,\footnote{For an experiment with \emph{Planck} resolution and noise, the degradation in $S/N$ when jointly fitting $\mathcal{A}_{\rm kSZ^2}$ and $b_g$ is about 15\% (see H16).} but this can also serve as a very useful consistency check, since the bias obtained must agree with that determined from external data (e.g., the galaxy auto-correlation).

The maximum $S/N$ ratio can be estimated by using Fisher's formula:
\be
\left( \frac{\Delta \mathcal{A}_{\rm kSZ^2}}{\mathcal{A}_{\rm kSZ^2}} \right)^{-2} = f_{\rm sky} \sum_\ell  \frac{  \left( 2 \ell + 1 \right) \left( C_\ell^{\rm kSZ^2 \times \delta_g}\right)^2}{C_\ell^{\bar{T}^2 \bar{T}^2 , f} C_\ell^{\delta_g \delta_g} +  \left( C_\ell^{\rm kSZ^2 \times \delta_g}\right)^2 }
\ee
where $f_{\rm sky}$ is the observed sky fraction, $C_\ell^{\delta_g \delta_g}$ is the tracer density power spectrum (including shot noise), and for $C_\ell^{\bar{T}^2 \bar{T}^2 , f} $ we use the Gaussian approximation:
\be
C_\ell^{\bar{T}^2 \bar{T}^2 , f} \approx 2 \int \frac{d^2 \mathbf{L}}{(2 \pi)^2} C_L^{ \bar{T}  \bar{T},f} C_{| \bl - \mathbf{L} |}^{ \bar{T} \bar{T},f} \,.
\ee
Here $C_\ell^{\bar{T}  \bar{T},f} = F^2(\ell) b^2(\ell) (C_\ell^{TT} + C_\ell^{\rm kSZ} + N_\ell)$ and $C_{\ell}^{TT}$ is the lensed primary CMB temperature power spectrum. The noise power spectrum $N_\ell$ is given by 
\be
N_\ell = \Delta^2_T b^{-2}(\ell) \approx \Delta^2_T \exp{\left(\frac{\theta^2_{\rm FWHM} \ell^2}{8 \ln2} \right)}
\ee
where $\Delta^2_T$ is the pixel noise level of the experiment (usually quoted in $\mu$K-arcmin) and $\theta_{\rm FWHM}$ is the beam full-width at half-maximum (FWHM).

Since $\mathcal{A}_{\rm kSZ^2} \propto f_{\rm free}^2$, if we are interested in a measurement of the free electron fraction $f_{\rm free}$, the fractional error is given by
\be
\left( \frac{\Delta f_{\rm free}}{f_{\rm free}} \right) \approx \frac12 \left( \frac{\Delta \mathcal{A}_{\rm kSZ^2}}{\mathcal{A}_{\rm kSZ^2}} \right) \,.
\ee

Table \ref{tab:forecast} shows the expected results for a selection of CMB experiments and large-scale structure probes. For concreteness we have picked the \emph{WISE} galaxy catalog and a catalog from the proposed \emph{SPHEREx}~\cite{2014arXiv1412.4872D} space-based experiment as our large-scale structure surveys of choice, but we note that the next decade will see a large number of galaxy surveys, both ground and space-based.  Details about the surveys considered here are given in Appendix~A.

\begin{figure}[ht]
\centering
\includegraphics[width=0.5\textwidth]{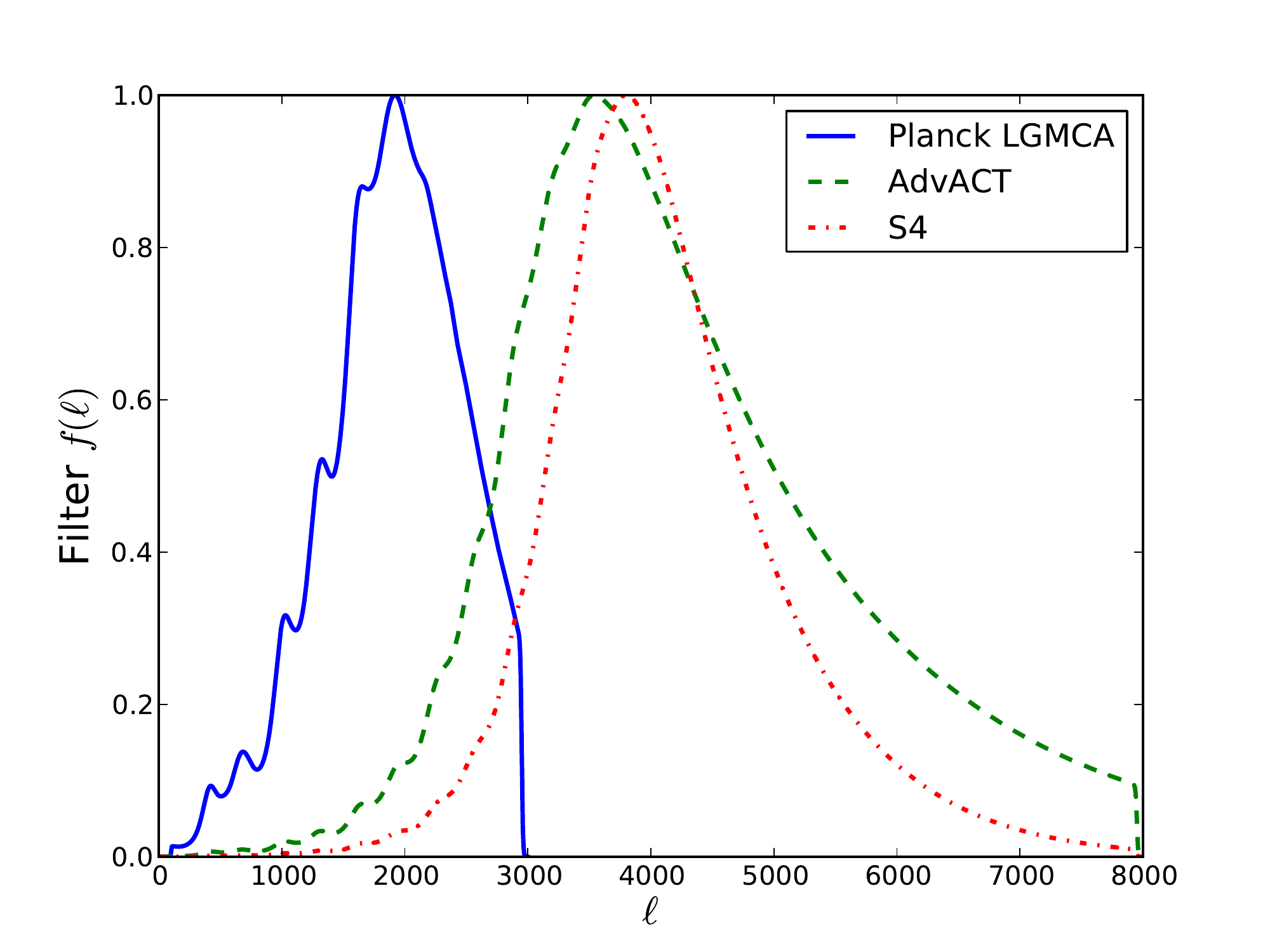}
\caption{Filters $f(\ell) = F(\ell)b(\ell)$ for the three CMB experiments considered here. The normalization is arbitrary and the results are independent of the normalization. The filter for a hypothetical \emph{CMB-S4} experiment is matched to the configuration of case 1 in Table \ref{tab:forecast_specs}.}
\label{fig.filters}
\end{figure}

The effective noise level for \emph{Advanced ACTPol}~\cite{Henderson2015} is determined by assuming that the component separation procedure yields a multiplicative increase over the proposed 150 GHz channel noise equal to that found for the 2015 \emph{Planck} + WMAP9 LGMCA map compared to the \emph{Planck} 143 GHz channel noise (a factor of $47 / 33 = 1.4$). The filters used in these forecasts are shown in Figure \ref{fig.filters}, $f(\ell) = F(\ell) b(\ell)$, where $F(\ell)$ is constructed from Equation~\ref{eq.Felldef} and $b(\ell)$ is the beam.

As seen in Table \ref{tab:forecast}, the statistical $S/N$ for future CMB experiments is enormous, and thus the actual results are likely to be limited by systematics such as foreground component separation or theoretical modeling uncertainties. These and other challenges are discussed in Section \ref{sec:challenges}.

\begin{table}[h]
\begin{center}
  \begin{tabular}{| c | c | c |}
    \hline 
     \ CMB experiment & beam FWHM  & effective noise\footnote{Here by ``effective noise'' we mean the residual cleaned CMB map noise after component separation.} \\
      \ & [arcmin] & $\Delta_T$ [$\mu$K-arcmin] \\ \hline \hline
    \emph{Planck} (2015 LGMCA map) & 5 & 47  \\ \hline
    \emph{Advanced ACTPol}  & 1.4 & 10 \\ \hline
    \emph{CMB-S4} (case 1) \footnote{Specifications for a future S4 experiment are not yet set, therefore here we consider a few cases for illustration purposes. Actual properties may be different.}  & 3 & 3 \\ \hline
    \emph{CMB-S4} (case 2) & 1	& 3 \\ \hline
     \emph{CMB-S4} (case 3) & 3	 & 1 \\ \hline
    \emph{CMB-S4} (case 4) & 1	& 1 \\ \hline
  \end{tabular}
  \caption{Specifications for the CMB experiments assumed in the forecasts.}
  \label{tab:forecast_specs}
\end{center}
\end{table}

\begin{table}[h]
\begin{center}
  \begin{tabular}{| c | c | c | c |}
    \hline
     \ &\ $f_{\rm sky}$\ & $\ell$ range & $\left( \frac{\Delta f_{\rm free}} {f_{\rm free}}\right)^{-1}$\\ \hline \hline
    \emph{Planck} $\times$ \emph{WISE} & 0.7 & 100 - 3000 & 5.2  \\ 
    \emph{Planck} $\times$ \emph{SPHEREx} & 0.7 & 100 - 3000 & 5.4 \\ \hline
    \emph{Advanced ACTPol} $\times$ \emph{WISE} & 0.5 & 100 - 8000 & 232 \\ 
    \emph{Advanced ACTPol} $\times$ \emph{SPHEREx} & 0.5 & 100 - 8000 & 280 \\ \hline
    \emph{CMB-S4} (case 1)  $\times$ \emph{WISE} & 0.5 & 100 - 8000 & 296 \\
    \emph{CMB-S4} (case 1) $\times$ \emph{SPHEREx} & 0.5 & 100 - 8000 & 356 \\ \hline
    \emph{CMB-S4} (case 2)  $\times$ \emph{WISE} & 0.5 & 100 - 8000 & 704 \\
    \emph{CMB-S4} (case 2) $\times$ \emph{SPHEREx} & 0.5 & 100 - 8000 & 866 \\ \hline
    \emph{CMB-S4} (case 3)  $\times$ \emph{WISE} & 0.5 & 100 - 8000 & 702 \\
    \emph{CMB-S4} (case 3) $\times$ \emph{SPHEREx} & 0.5 & 100 - 8000 & 858 \\ \hline
    \emph{CMB-S4} (case 4)  $\times$ \emph{WISE} & 0.5 & 100 - 8000 & 822 \\
    \emph{CMB-S4} (case 4) $\times$ \emph{SPHEREx} & 0.5 & 100 - 8000 & 1014 \\

    \hline
  \end{tabular}
  \caption{Forecasts for determining $f_{\rm free}$ from the kSZ$^2$--galaxy cross-correlation. The baryon profile on small scales is very uncertain and in order to minimize the theoretical uncertainties, we have fixed the filter for all S4 cases to the lower resolution case 1. In all cases the noise and the resolution are treated self-consistently.}
  \label{tab:forecast}
\end{center}
\end{table}

\begin{figure}[ht]
\centering
\includegraphics[width=0.47\textwidth]{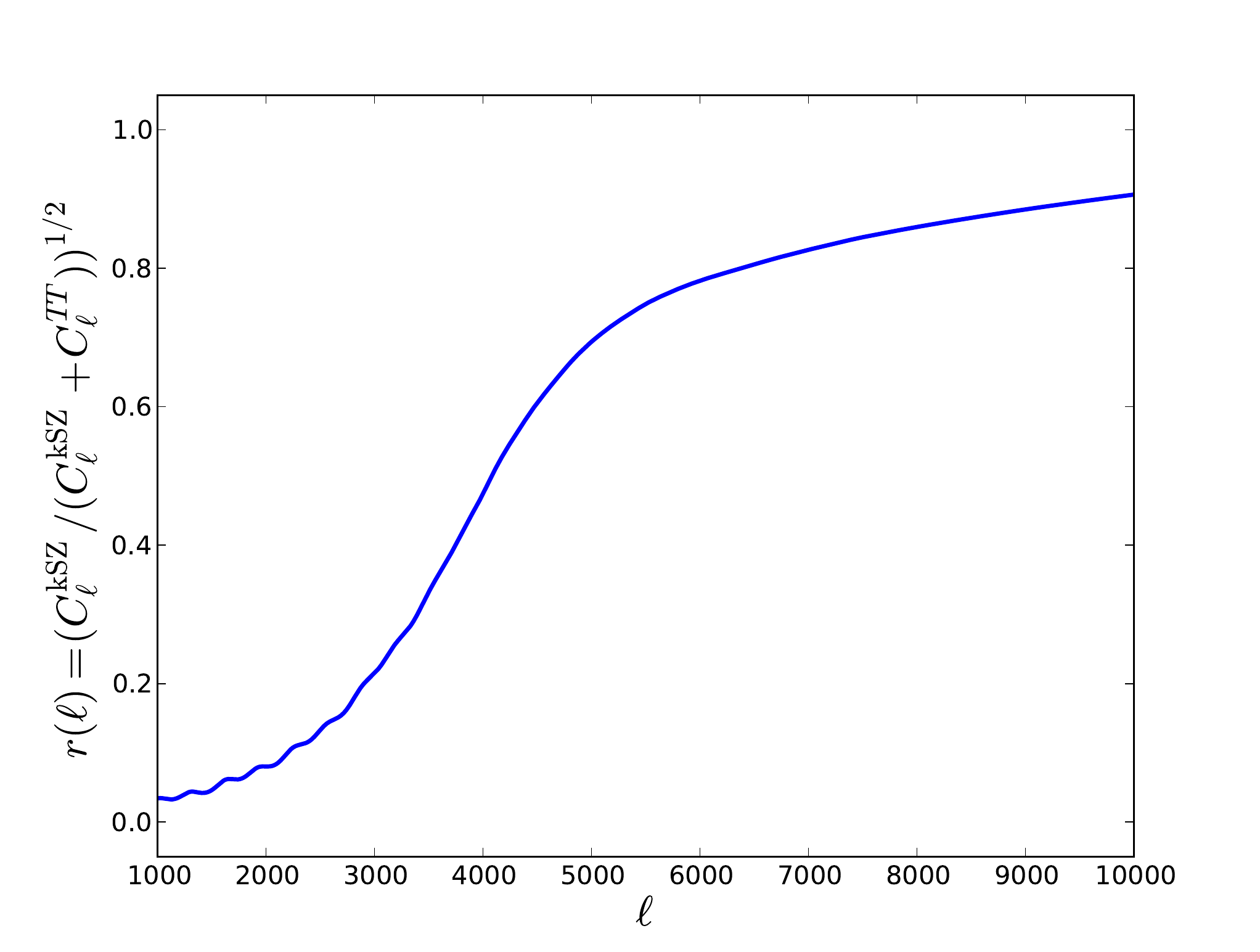}
\caption{Correlation coefficient between the kSZ ``field'' and the total temperature fluctuation field that has a blackbody frequency spectrum. Noise is not included here since it is experiment-dependent. The ``kSZ'' field includes both late-time and reionization contributions. Note that while the exact shape and amplitude of the kSZ signal are still uncertain, the qualitative features should be correct. Here the $C_\ell^{TT}$ power includes lensing.}
\label{fig.corr_coeff}
\end{figure}

For a CMB experiment with the angular resolution of \emph{Planck}, this method is suboptimal (in terms of $S/N$ per object) when 3D information is available and should only be used in the absence of reliable spectroscopic redshifts.  This is easy to understand: our method uses the observed CMB temperature as a proxy for the cluster peculiar velocity, rather than the 3D position of the tracers.  On large angular scales ($\ell \lesssim 3000$), the primary anisotropy is much larger than the kSZ amplitude and the signal-to-noise per mode is very small. As high-resolution CMB experiments allow us to access smaller scales, we expect very high-$S/N$ detections with \emph{Advanced ACTPol} and \emph{CMB-S4}. In fact, at $\ell \gtrsim 4000$, the fluctuation field is dominated by kSZ and not by the primary anisotropy. This point is illustrated in Figure \ref{fig.corr_coeff}, where we show the correlation coefficient between the total temperature field that has a blackbody spectrum (i.e., lensed primary CMB and kSZ on the scales of interest) and the kSZ field.  While this cross-correlation is small at low $\ell$ (including all of the $\ell$ range probed by \emph{Planck} and \emph{WMAP}), it grows to order unity at high $\ell$. This means that in the absence of other frequency-dependent foregrounds and noise, high-resolution CMB maps are a direct probe of the integrated electron momentum.

Also note that even at \emph{Planck} resolution, this method allows us to use much larger photometric catalogs such as \emph{WISE}, instead of smaller spectroscopic samples. As seen in Table \ref{tab:forecast}, we expect the combination of \emph{Planck} and \emph{WISE} to yield constraints that are comparable to recent analyses that use the full 3D (spectroscopic) information in the galaxy density field, thanks to the fact that we can use a much larger sample of tracer objects ($\sim 10^8$ in our work with \emph{WISE}, compared to $\sim 10^4 - 10^5$ for previous works \cite{2015arXiv151006442S, 2015arXiv150403339P, Handetal2012}).

\section{CMB Lensing Contribution}
\label{sec:lensing}
Since our kSZ${}^2$ estimator is quadratic in the CMB temperature, it can potentially receive a contribution from weak lensing of the CMB, due to matter inhomogeneities between us and the surface of last scattering (see Ref.~\cite{2006PhR...429....1L} for a review on CMB lensing).  In this section, we define $\T = \Delta T /T$  to be the unlensed (primary) CMB temperature fluctuation and $\Tt$ be the corresponding lensed fluctuation. 

We first note that if we could observe the CMB with an infinitesimally small beam and did not apply any filter, then the lensing contribution to our estimator would vanish. This is because CMB lensing preserves the total variance, since the lensing amounts to a remapping of perturbations on the last scattering surface to a slightly different point in the sky \cite{2006PhR...429....1L}.

This argument no longer applies when we observe the CMB through a finite resolution experiment and the map is filtered as described above; in this case the weak lensing contribution can be large.  As before, we define the lensed $\Tf(\l) = f(\ell) \Tt(\l)$, where $f(\ell) = F(\ell) b(\ell)$ is the product of a filter $F$ and the beam function $b$. We would like to compute the Fourier transform of $\langle \Tf^2(\x) \delta_g (\y) \rangle$:
\begin{align}
& \langle \Tf^2(\l_1) \delta_g (\l_2) \rangle = \nn \\
& =  \int \frac{d^2 \L}{(2 \pi)^2} \langle \Tf(\L) \Tf(\l_1- \L)  \delta_g (\l_2) \rangle \nn  \\
& =  \int \frac{d^2 \L}{(2 \pi)^2} f(L) f(|\l_1 - \L|) \langle \Tt(\L) \Tt(\l_1- \L)  \delta_g (\l_2) \rangle \,.
\label{eq:first_lens}
\end{align}

The lensed fluctuation field can be expanded in terms of the unlensed field \cite{2006PhR...429....1L}:
\be
\Tt(\x) = \T(\x) + \nabla \psi \cdot \nabla \T(\x) + \ldots
\ee
where $\psi$ is the lensing potential, so that we can express 
\be
[\nabla \psi \cdot \nabla \T](\L) = - \int \frac{d^2 \L'}{(2 \pi)^2} \L' \cdot (\L - \L') \psi(\L') \T(\L - \L') \,.
\ee
Up to first order in the lensing potential we have
\begin{align}
& \langle \Tt(\L) \Tt(\l_1- \L) \delta_g (\l_2) \rangle = \langle \T(\L) \T(\l_1- \L)  \delta_g (\l_2) \rangle + \nn \\
& \langle [\nabla \psi \cdot \nabla \T](\L) \Tt(\l_1- \L)  \delta_g (\l_2) \rangle + (\L \rightarrow \l_1- \L) + \ldots
\label{eq:second_lens}
\end{align}
The first term is simply the fiducial $C_\ell^{\rm{kSZ}^2 \times \delta_g}$ for the kSZ$^2$-galaxy cross-correlation that was computed in Section~\ref{sec:theory}, while the second and third terms are the lowest order CMB lensing contribution and are equal in magnitude by symmetry. Plugging Equation \ref{eq:second_lens} into \ref{eq:first_lens} we find
\begin{widetext}
\be
\displaystyle
\langle \Tf^2(\l_1) \delta_g (\l_2) \rangle = \langle \T_f^2(\l_1) \delta_g (\l_2) \rangle - 2  \int \frac{d^2 \L}{(2 \pi)^2} f(L) f(|\l_1 - \L|) \int \frac{d^2 \L'}{(2 \pi)^2} \L' \cdot (\L - \L') \langle \psi(\L') \T(\L - \L') \T(\l_1- \L)  \delta_g (\l_2) \rangle + \ldots
\label{eq:third_lens}
\ee
\end{widetext}

The four-point function of the form $\langle \psi \Theta \Theta \delta_g \rangle$ on the right-hand side of Equation \ref{eq:third_lens} can be decomposed into a connected four-point function (technically non-vanishing because of ISW, but subdominant to the other terms in the range of scales considered here), and two non-zero contractions $\langle \psi \delta_g \rangle \langle \T \T \rangle$ and $\langle \psi \T \rangle \langle \T \delta_g \rangle$, the latter again non-zero due to ISW. Consider the first one and write:
\ba
\langle \psi(\L') \delta_g(\l_2) \rangle &=& (2\pi)^2 \ C^{\psi \delta_g}_{\ell_2} \ \delta_D(\L' + \l_2) \nn \\
\langle \T(\L - \L') \T(\l_1- \L) \rangle &=& (2\pi)^2 \ C^{TT}_{|\l_1 - \L|} \ \delta_D(\l_1 - \L') \nn
\ea
Then the main correction due to lensing\footnote{Here $C_{\ell}^{TT}$ denotes the \emph{unlensed} primary anisotropy power spectrum.} is (from the right-hand side of Equation~\ref{eq:third_lens})
\be
-2  \int \frac{d^2 \L}{(2 \pi)^2} f(L) f(| \L - \l_1|) \ \l_1 \cdot (\L-\l_1) \ C^{\psi \delta_g}_{\ell_1} \ C^{TT}_{|\L -\l_1 |} \,.
\label{eq:lens_leak}
\ee
Similarly, the other contraction gives rise to 
\be
-2  \int \frac{d^2 \L}{(2 \pi)^2} f(L) f(| \L - \l_1|) \ \l_1 \cdot (\L-\l_1) \ C^{\psi T}_{\ell_1} \ C^{T \delta_g}_{|\L -\l_1 |}
\ee
which is due to ISW and numerically is found to be factor of $\sim 10^4 - 10^5$ smaller than the former contribution on the scales considered here. Thus it will be neglected in the following.

Changing variables in Equation~\ref{eq:lens_leak} to $\L' = \L - \l_1$, we can rewrite the leading-order lensing contribution as
\begin{align}
\Delta C_{\ell}^{T^2 \times \delta_g} \approx -2 \frac{\ell \ C_{\ell}^{\psi \delta_g}}{(2 \pi)^2} & \displaystyle\int_{0}^{\infty} dL' \ L'^2 f(L') C_{L'}^{TT}  \nn \\
&\displaystyle\int_0^{2 \pi} d \phi \ f(|\L' +\l |) \cos \phi 
\label{eq:lensleak}
\end{align}
Finally, we see that in the absence of a filter and beam (i.e., $f(\ell) = $ constant), the lensing correction vanishes as expected. Examples of the lensing contribution are shown in Figures \ref{fig.lens_leakage} and \ref{fig.kSZ2xWISEdata}.  It displays a characteristic oscillatory behavior that makes it nearly orthogonal to the kSZ${}^2$ signal.

Heuristically, we interpret the shape of the lensing contribution as follows.  The overall effect of lensing is to slightly shift the amount of power that lies within the filter applied in our analysis, i.e., to slightly change the local variance in the filtered temperature map.  There are two competing effects due to lensing.  First, in overdense (underdense) regions, lensing magnification (demagnification) shifts the temperature power spectrum to lower (higher) multipoles, thus decreasing (increasing) the amount of power within our filter.  Since the large-scale structure tracer density will fluctuate higher (lower) in overdense (underdense) regions, this effect produces a negative correlation between the local variance of the filtered map and the tracer density map.  Second, lensing transfers temperature power from low to high multipoles, thus increasing the amount of power within our filter in regions with strong density fluctuations.  This effect produces a positive correlation between the local variance of the filtered map and the tracer density map.  The oscillatory shape of the overall lensing contribution comes from the interplay of these two effects: our results indicate that the first effect dominates on large scales, while the second effect dominates on small scales, with a zero-crossing at $\ell \approx 1600$--$1700$ for our \emph{Planck}/\emph{WMAP}/\emph{WISE} analysis (see Figure~\ref{fig.kSZ2xWISEdata}).  The exact magnitude and shape of the lensing contribution depends on the CMB experiment, $\ell$-space filter, and large-scale structure survey used in the analysis.

\section{Comparison to Numerical Simulations}
\label{sec:sims}
In this section we compare our theoretical predictions for the kSZ$^2$ signal and lensing contribution to two different sets of numerical simulations.  The first is a cosmological hydrodynamics simulation \cite{BBPSS2010}, while the second is constructed from a dark-matter-only tree-particle-mesh simulation, in which halos are populated with gas in post-processing using a polytropic equation of state and hydrostatic equilibrium \cite{2010ApJ...709..920S}. In this section only, the cosmological parameters for the theory curves are chosen to match the respective simulations and will in general differ from the fiducial cosmology assumed in the rest of the paper.

As a first test, we set the filter $f(\ell)$ to a constant and compare our theoretical prediction to the simulations from \cite{BBPSS2010}. These are hydrodynamic simulations of cosmological volumes (box side-length $L=165$ Mpc$/h$) using a modified version of the GADGET-2 code \cite{2005MNRAS.364.1105S}. 
Included in these simulations are sub-grid physics models for active galactic nuclei (AGN) feedback \cite{BBPSS2010}, cosmic ray physics
\cite{2006MNRAS.367..113P, 2007A&A...473...41E, 2008A&A...481...33J}, radiative cooling, star formation, galactic winds, and supernova feedback
\cite{2003MNRAS.339..289S}. 
The halo catalogs from these simulations are incomplete below masses of $\approx 5 \times 10^{13} \, M_{\odot}$~\cite{BBPS2012b}, and thus we cannot construct simulated galaxy density maps to mock the \emph{WISE} or \emph{SPHEREx} samples.  Instead, we consider weak gravitational lensing convergence ($\kappa_{\rm CFHT}$) as the large-scale structure tracer of choice in this analysis.  For the present comparison, we construct mock lensing convergence maps using mass shells extracted from the simulations and a source galaxy redshift distribution matching that of the Canada-France-Hawaii Telescope Lensing Survey (CFHTLenS)~\cite{Heymansetal2012}.

\begin{figure}[ht]
\centering
\includegraphics[width=0.5\textwidth]{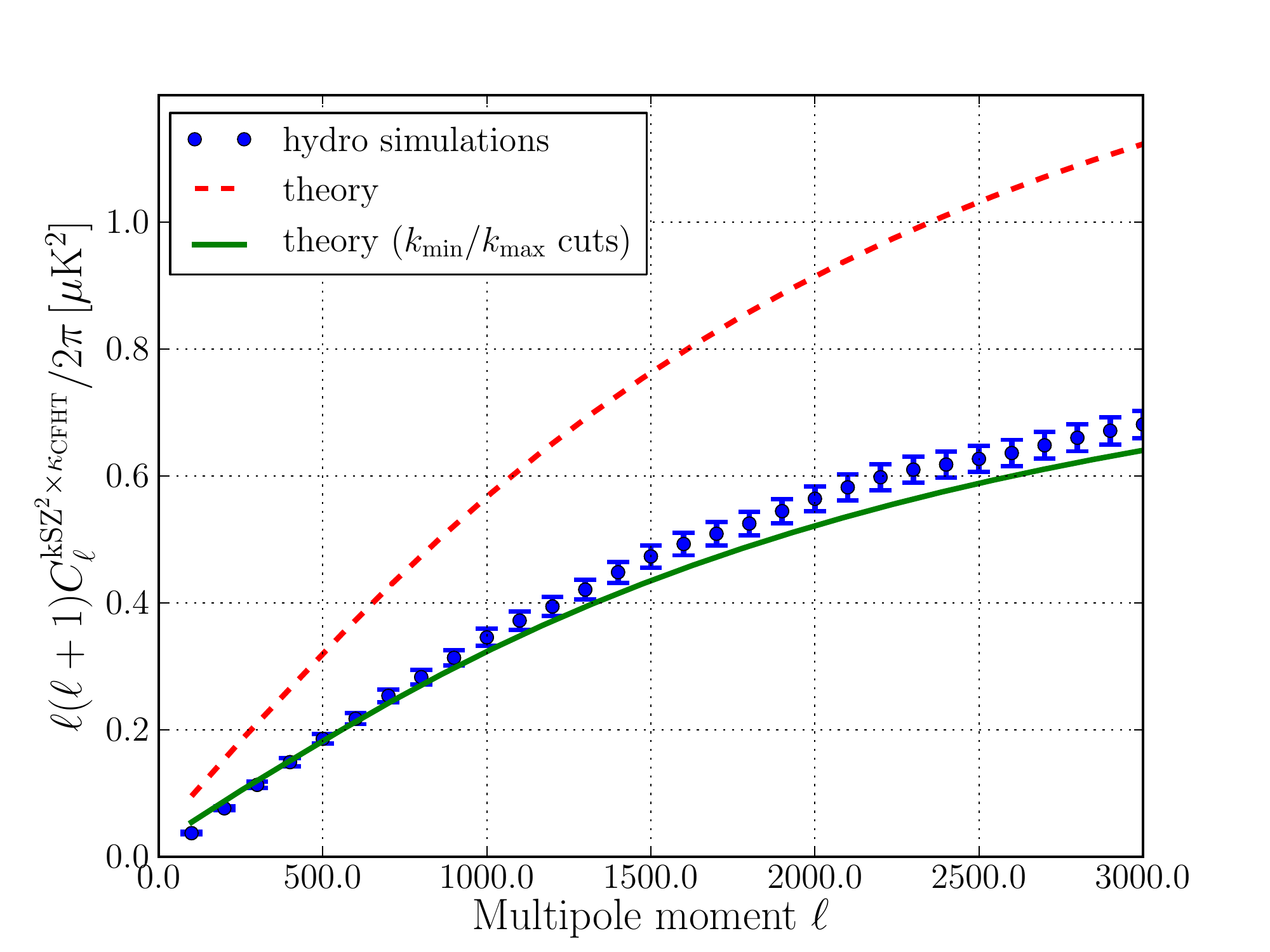}
\caption{Theoretical cross-correlation between the unfiltered kSZ${}^2$ signal and CFHTLenS-like weak lensing convergence maps. The blue points with error bars show the result measured from cosmological hydrodynamics simulations \cite{BBPSS2010}. The dashed red curve shows the fiducial theory computation, while the solid green curve shows the theory computation with wavenumber cut-offs matching those of the simulation ($k_{\rm min} = 0.038\ h/$Mpc and $k_{\rm max} = 76\ h/$Mpc). This comparison shows that the simulation results are biased low due to the lack of super-box long wavelength velocity modes. This effect can be large, as seen here, but if properly accounted for, the theory and simulations agree to $\lesssim 5$\% over the whole range considered. Here we have used $f_{\rm free} = 0.85$, independent of redshift. The small difference between theory and simulations might be explained by the redshift evolution of $f_{\rm free}$ or by the intrinsic uncertainty on our theory curve due to the fitting function for the non-linear bispectrum, which is of order 5-10\% \cite{GilMarin:2011ik}.}
\label{fig.kSZ2xkappa_unfiltered}
\end{figure}

The kSZ and the CHFTLenS-like lensing convergence maps are made at each redshift snapshot following the methods described in \cite{BBPS2012b} and \cite{BHM2014}, respectively. We compute the cross-power spectrum for each redshift output and then average the cross-power spectra over ten initial condition realizations. We sum these average spectra over the redshift outputs to compute the final spectrum. The results of this comparison are shown in Figure \ref{fig.kSZ2xkappa_unfiltered}, with error bars computed from the scatter amongst the ten realizations.

One important caveat when comparing theory to simulations is that the velocity field is coherent on very large scales and thus finite-box simulations can underpredict the expected signal, since they lack contributions from velocity modes with wavelength larger than the box size \cite{2013ApJ...769...93P}. To be more quantitative, from Equation \ref{bispectrum}, the signal is proportional to $v^2_{\rm rms}$, and we find that about half of the contribution to $v^2_{\rm rms}$ comes from $k < 0.06\ h/$Mpc. As seen in Figure \ref{fig.kSZ2xkappa_unfiltered}, the agreement between theory and simulations is excellent when using the same $k_{\rm min} $ and $k_{\rm max}$ as the simulations, but there is a large discrepancy if we neglect the effect of the finite box size.

\begin{figure}[ht]
\centering
\includegraphics[width=0.5\textwidth]{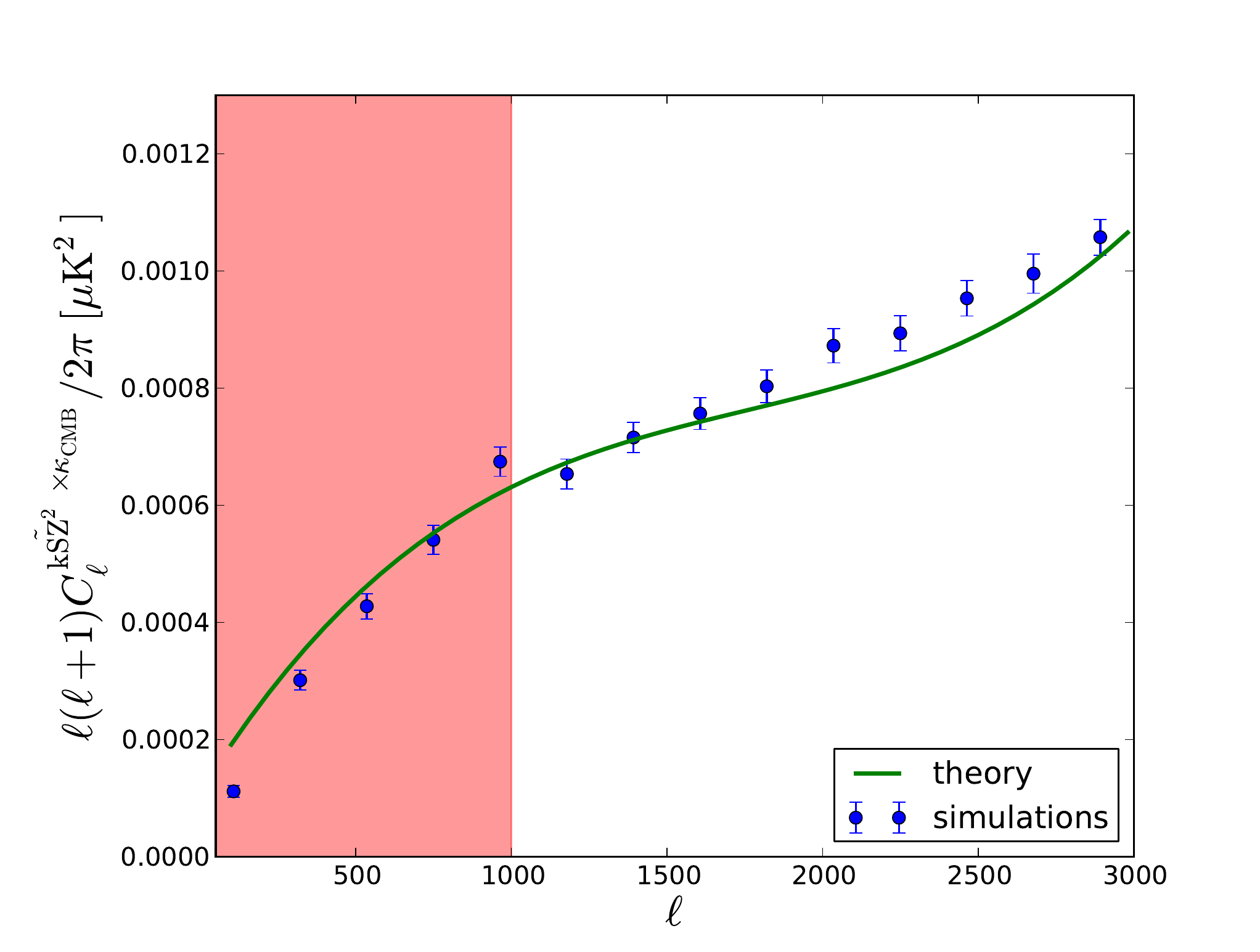}
\caption{Cross-correlation between the filtered kSZ${}^2$ signal and CMB lensing convergence maps.  The blue points with error bars show the result measured from the simulation of Ref.~\cite{2010ApJ...709..920S}, using a filter $f(\ell)$ appropriate for \emph{Planck} data. The agreement is better than 10\% at high $\ell$ and the difference at very low $\ell$ is likely due to the fact that these simulations do not accurately predict the kSZ power spectrum on large scales (see Section 2.4 of~\cite{2010ApJ...709..920S}). The shaded region at $\ell < 1000$ represents the scales that may be unreliable in the simulation.} 
\label{fig.kSZ2xkappa_filtered}
\end{figure}

Next we compare our predictions to the full-sky simulation of Ref.~\cite{2010ApJ...709..920S}, with a non-trivial filter $f(\ell)$ that includes the weighting and beam appropriate for the \emph{Planck} experiment (in particular, as constructed from the 2013 LGMCA map~\cite{Bobinetal2014}).  Note that the simulation box is $1$~Gpc$/h$ on a side, so effects related to the low-$k$ cut-off discussed above are substantially reduced here.  For this analysis, we consider CMB lensing convergence ($\kappa_{\rm CMB}$) as our large-scale structure tracer, since ray-traced maps of this quantity have already been computed from this simulation. The results are shown in Figure \ref{fig.kSZ2xkappa_filtered}.  In this case, since only one simulation is available, we estimate error bars from the scatter within each multipole bin.

We find agreement between theory and simulation to better than 10\% at $\ell \gtrsim 500$. There is a minor discrepancy at very low $\ell$, but this might be explained by the filtering applied to the simulations: because of the way that the lightcone was constructed, the kSZ signal from the inter-galactic medium was overpredicted on large scales and therefore a filter of the form $w(\ell) = 1 - e^{-(\ell/500)^2}$ was applied to the kSZ map to suppress the large-scale excess. The authors of \cite{2010ApJ...709..920S} caution that ``since the simple filtering modifies the signal at $\ell < 1000$, the maps should not be used to predict the kSZ signal at these scales'' and thus the slight low-$\ell$ discrepancy is not a significant cause for concern.

\begin{figure}[ht]
\centering
\includegraphics[width=0.5\textwidth]{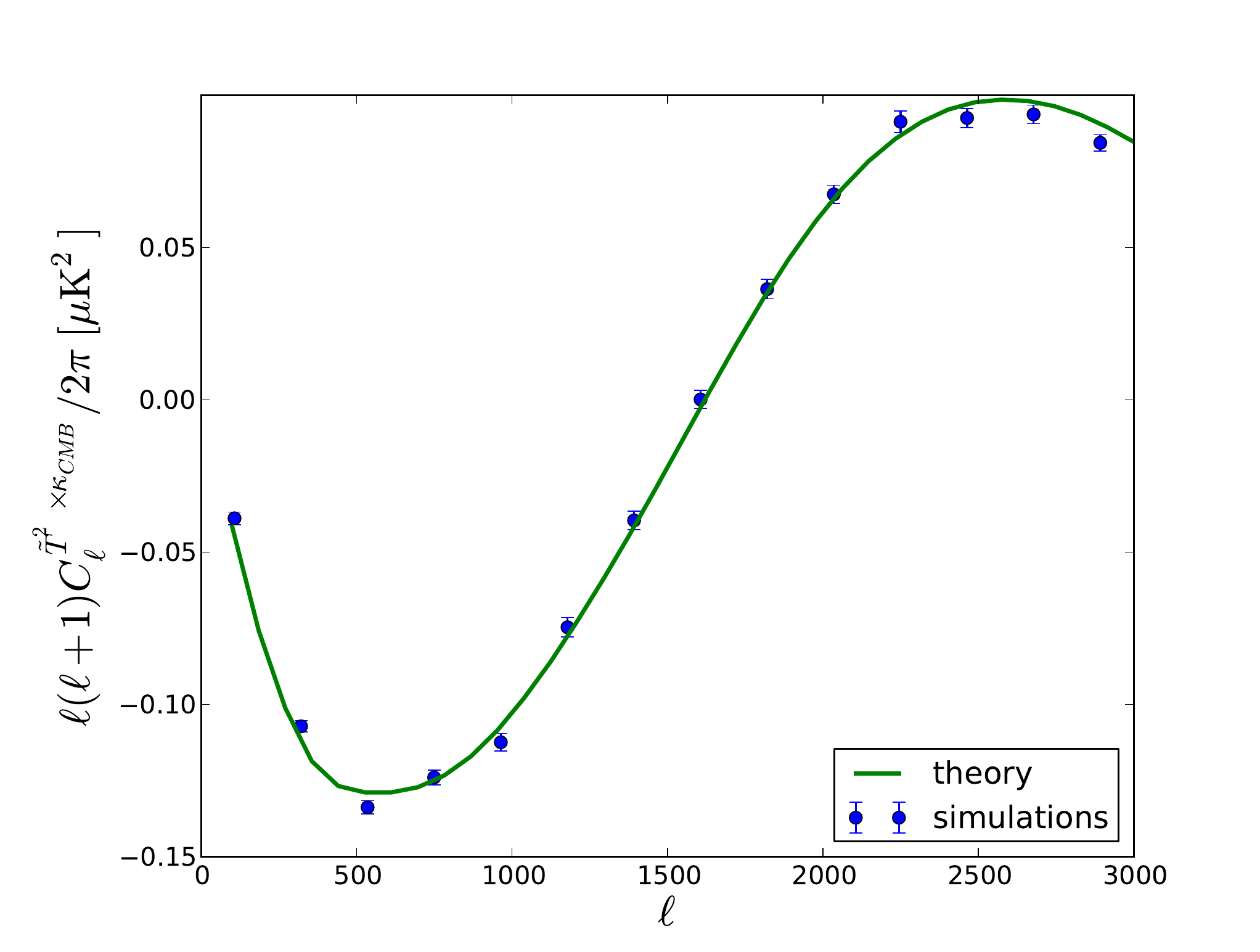}
\caption{Lensing leakage in the kSZ${}^2$ estimator, inferred from simulated lensed CMB temperature maps cross-correlated with CMB lensing convergence maps, i.e., $\langle \tilde{T}_f^2 \ \kappa_{\rm CMB} \rangle$.  (To be clear, $\kappa_{\rm CMB}$ is simply the large-scale structure tracer considered in this test; the lensing leakage calculation can be applied to any tracer.) The blue points show the cross-correlation measured from the simulation of Ref.~\cite{2010ApJ...709..920S}, while the solid line is the analytic calculation presented in this paper (see Equation~\ref{eq:lensleak}), and is obtained using fitting functions for the non-linear matter power spectrum of \cite{2012ApJ...761..152T}. The agreement is better than 3\% over the entire multipole range probed here, with the exception of the highest $\ell$ point (where the non-linear corrections are largest and which deviates from the theory curve by $\approx$6\%).}
\label{fig.lens_leakage}
\end{figure}

Finally, we test our lensing leakage prediction from Equation \ref{eq:lensleak}, using $\kappa_{\rm CMB}$ rather than $\delta_g$ as the large-scale structure tracer of choice. For this comparison, we calculate the cross-correlation between the square of the lensed, filtered CMB temperature map (with no other secondary anisotropy) and the CMB weak lensing convergence map. The result is shown in Figure \ref{fig.lens_leakage}, indicating an agreement to better than 6\% on all scales. Therefore we conclude that higher order corrections are subleading and can be neglected at the current level of precision.

\section{Example: Measurement Using \emph{WMAP}, \emph{Planck}, and \emph{WISE}}
\label{sec:H16}
In H16, we recently presented the first measurement of the kSZ signal using this method. Here we briefly summarize the analysis as an example of an application to real data.  Some specific technical details are found in H16. We also discuss several of the challenges of this measurement in Section VII.

We use a cleaned CMB temperature map constructed from a joint analysis of the nine-year \emph{WMAP}~\cite{Bennettetal2013} and \emph{Planck} full mission~\cite{2015arXiv150201582P} full-sky temperature maps~\cite{Bobinetal2015}.\footnote{{\tt http://www.cosmostat.org/research/cmb/planck\char`_wpr2}}

The CMB is separated from other components in the microwave sky using ``local-generalized morphological component analysis'' (LGMCA), a technique relying on the sparse distribution of non-CMB foregrounds in the wavelet domain.  We refer the reader to~\cite{Bobinetal2013,Bobinetal2015} for a thorough description of this component separation technique and characterization of the resulting maps.  The method reconstructs a full-sky CMB map with minimal dust contamination and essentially zero contamination from the thermal SZ (tSZ) effect, which is explicitly projected out in the map construction (unlike in, e.g., the official \emph{Planck} SEVEM, NILC, or SMICA component-separated CMB maps, which all possess significant tSZ residuals). Since the kSZ signal preserves the CMB blackbody spectrum, it is not removed by the component separation algorithm.  We further clean the LGMCA map to explicitly deproject any residual emission associated with the \emph{WISE} galaxies (e.g., from dust) --- see H16 for details.

As discussed in Section II, a filter is applied to the CMB map before squaring in real space to downweight scales that are dominated by the primary CMB or noise.  The filter used in H16 is shown in Figure \ref{fig.filters} (including multiplication by the FWHM $=5$ arcmin beam of the LGMCA map).

The \emph{WISE} \cite{2010AJ....140.1868W} source catalog contains more than 500 million objects, roughly 70\% of which are star-forming galaxies \cite{2013AJ....145...55Y}. Color cuts can be used to separate galaxies from stars and other objects.  We use the same selection criteria as Ref.~\cite{Ferraroetal2014} to select a sample of galaxies, originally based on previous work \cite{2011ApJ...735..112J}, and we refer the reader to these papers for a detailed explanation. 

The redshift distribution of \emph{WISE}-selected galaxies has been shown to be fairly broad, with a peak at $z \approx 0.3$ and extending to $z = 1$ \cite{2013AJ....145...55Y}. Here we note that the galaxy selection is imperfect and that there is some residual stellar contamination, especially close to the Galactic plane. However, Galactic stars are expected to be uncorrelated with the kSZ signal, and any contamination will only lead to larger noise (which is taken into account in our analysis), but not a bias. For this reason, we apply a conservative mask that leaves $f_{\rm sky} = 0.447$ and 46.2 million galaxies.

The theory curve is the sum of the theoretical kSZ${}^2$ and lensing templates, the amplitude of each being $\mathcal{A}_{\rm kSZ^2} b_g$ and $b_g$, respectively (where we have defined $\mathcal{A}_{\rm kSZ^2} \propto f_{\rm free}^2$ as the amplitude of the kSZ$^2$ signal, with a fiducial expectation of unity). The best fit amplitude is found by minimizing the function
\be
\chi^2(\mathcal{A}_{\rm kSZ^2}, b_g) = (\mathbf{d} - \mathbf{t}(\mathcal{A}_{\rm kSZ^2}, b_g))^T C^{-1} (\mathbf{d} - \mathbf{t}(\mathcal{A}_{\rm kSZ^2}, b_g))
\ee
where the theory template is
\be
\mathbf{t}(\mathcal{A}_{\rm kSZ^2}, b_g) = \mathcal{A}_{\rm kSZ^2} b_g \ \mathbf{t}_{{\rm kSZ}^2} + b_g \ \mathbf{t}_{\rm lens} \,, 
\label{eq.thtemplate}
\ee
$\mathbf{d}$ is the data vector (from the measured cross-correlation) and $C^{-1}$ is the inverse of the noise covariance matrix estimated from the data itself, sourced by primary CMB fluctuations and other sources of noise.
For the best fit we find $\chi^2_{\rm b.f.} \ / \ {\rm dof}$ = 13.1 / 11, indicating a good fit.

\begin{figure}[ht]
\centering
\includegraphics[width=0.5\textwidth]{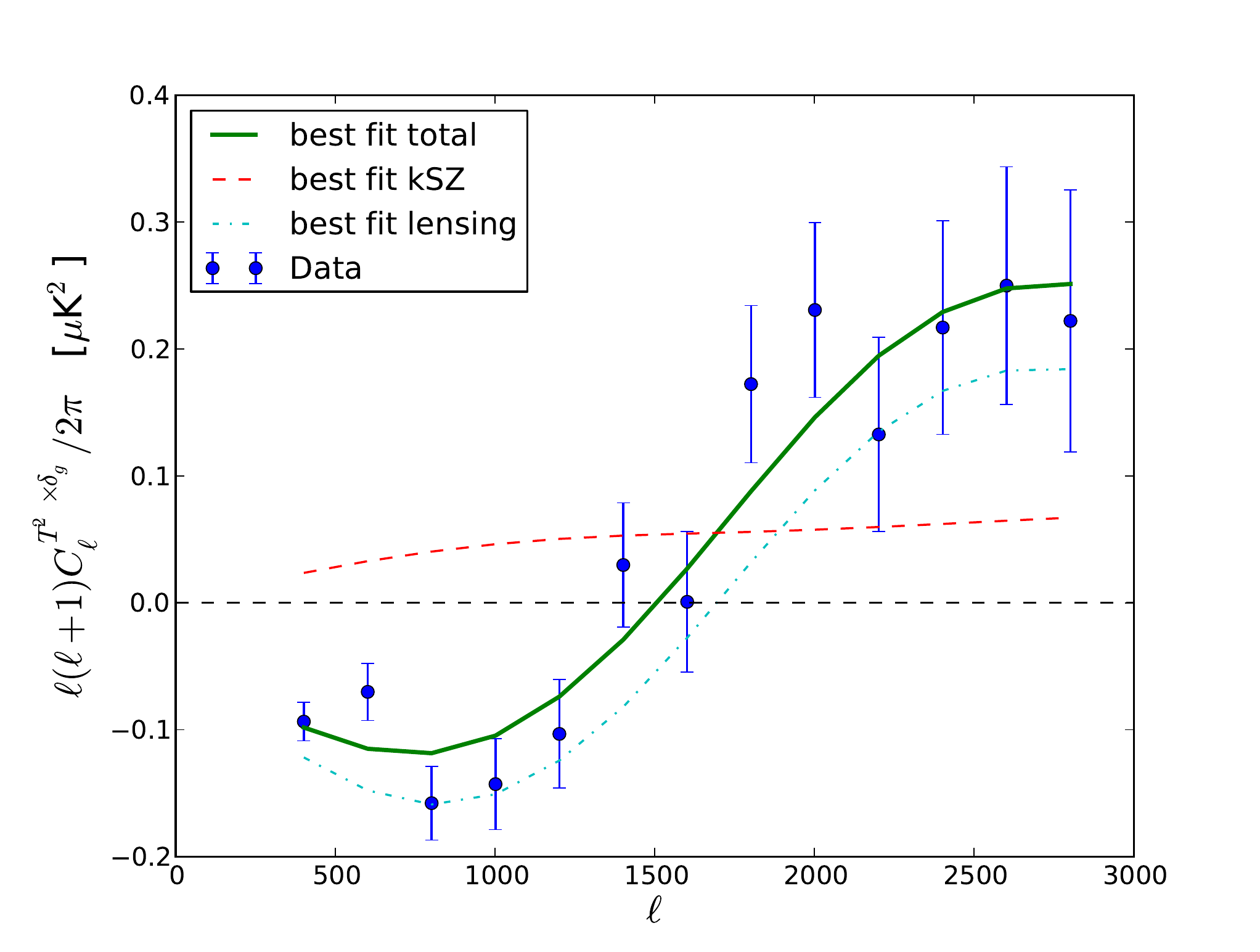}
\caption{Results for the $C_\ell^{\rm{kSZ}^2 \times \delta_g}$ analysis of H16, shown in blue. The dashed red curve is the best-fit kSZ${}^2$ template, the dash-dotted cyan curve is the best-fit lensing template, and the solid green curve is the sum of the two. No external prior on the galaxy bias is used in the fit shown in this plot.}
\label{fig.kSZ2xWISEdata}
\end{figure}

\begin{figure}[ht]
\centering
\includegraphics[width=0.5\textwidth]{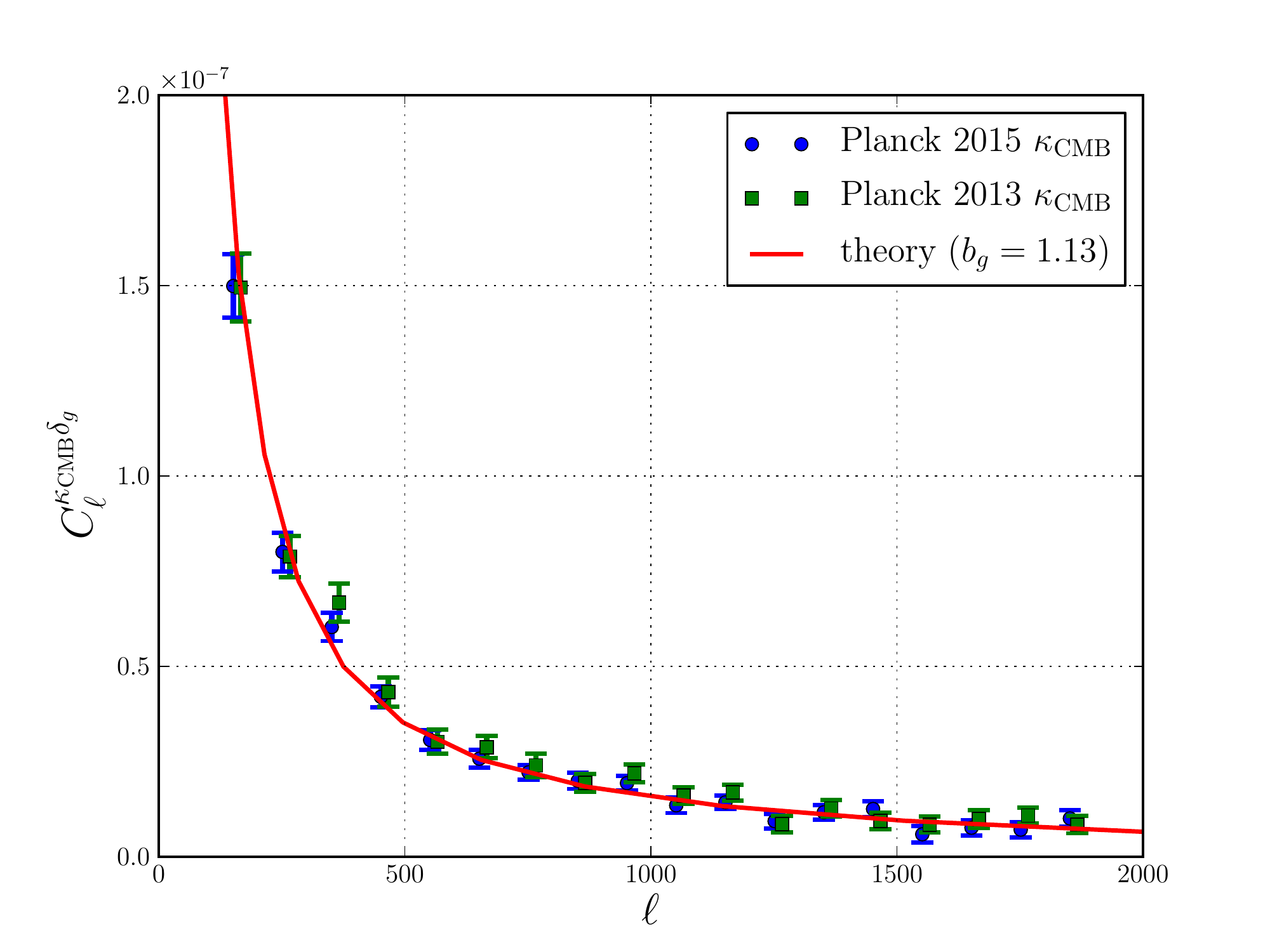}
\caption{Cross-correlation between Planck lensing convergence maps and WISE galaxies, shown for both the 2013 and 2015 version of the lensing maps. The best fit galaxy bias from the 2015 map is $b_g = 1.13 \pm 0.02$.}
\label{fig.planckappaXwise}
\end{figure}

Figure \ref{fig.kSZ2xWISEdata} shows the total best fit to the data, as well as the individual contributions from the kSZ$^2$ and lensing templates (matching Figure 1 of H16). In our fiducial analysis, we marginalize over the lensing contribution, but as a check we also obtain the galaxy bias by cross-correlating the \emph{WISE} sample with \emph{Planck} CMB lensing maps \cite{Planck2013lensing, Planck2015lensing}. This cross-correlation is shown in Figure \ref{fig.planckappaXwise}.

\begin{figure}[ht]
\centering
\includegraphics[width=0.5\textwidth]{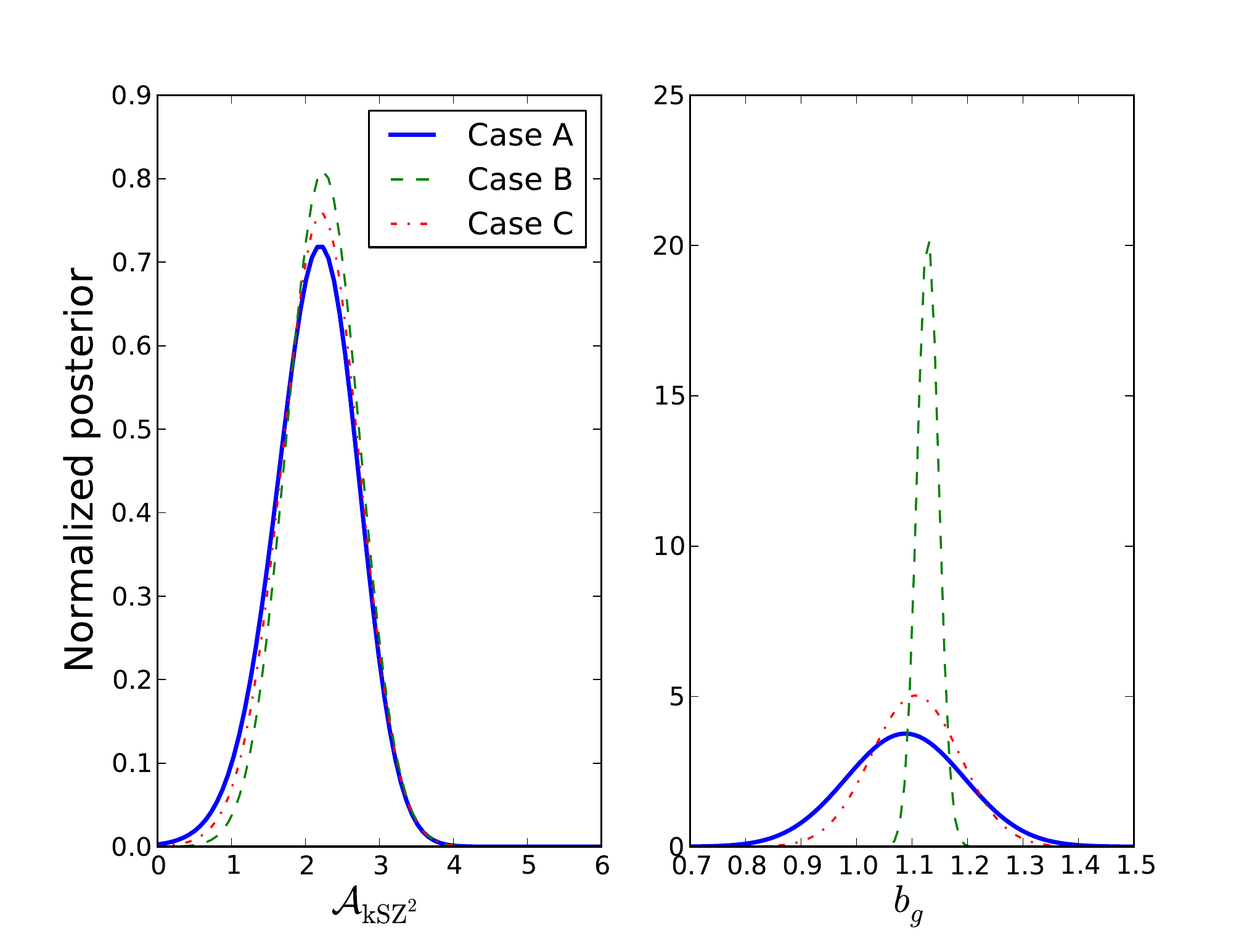}
\caption{Posterior distribution for $\mathcal{A}_{\rm kSZ^2}$ and  $b_g$ for the three analysis cases given in Table \ref{tab:results}. }
\label{fig.postAb}
\end{figure}

The posteriors for $\mathcal{A}_{\rm kSZ^2}$ with and without the prior on $b_g$ from the external CMB lensing data are shown in Figure \ref{fig.postAb}.

\begin{figure}[ht]
\centering
\includegraphics[width=0.5\textwidth]{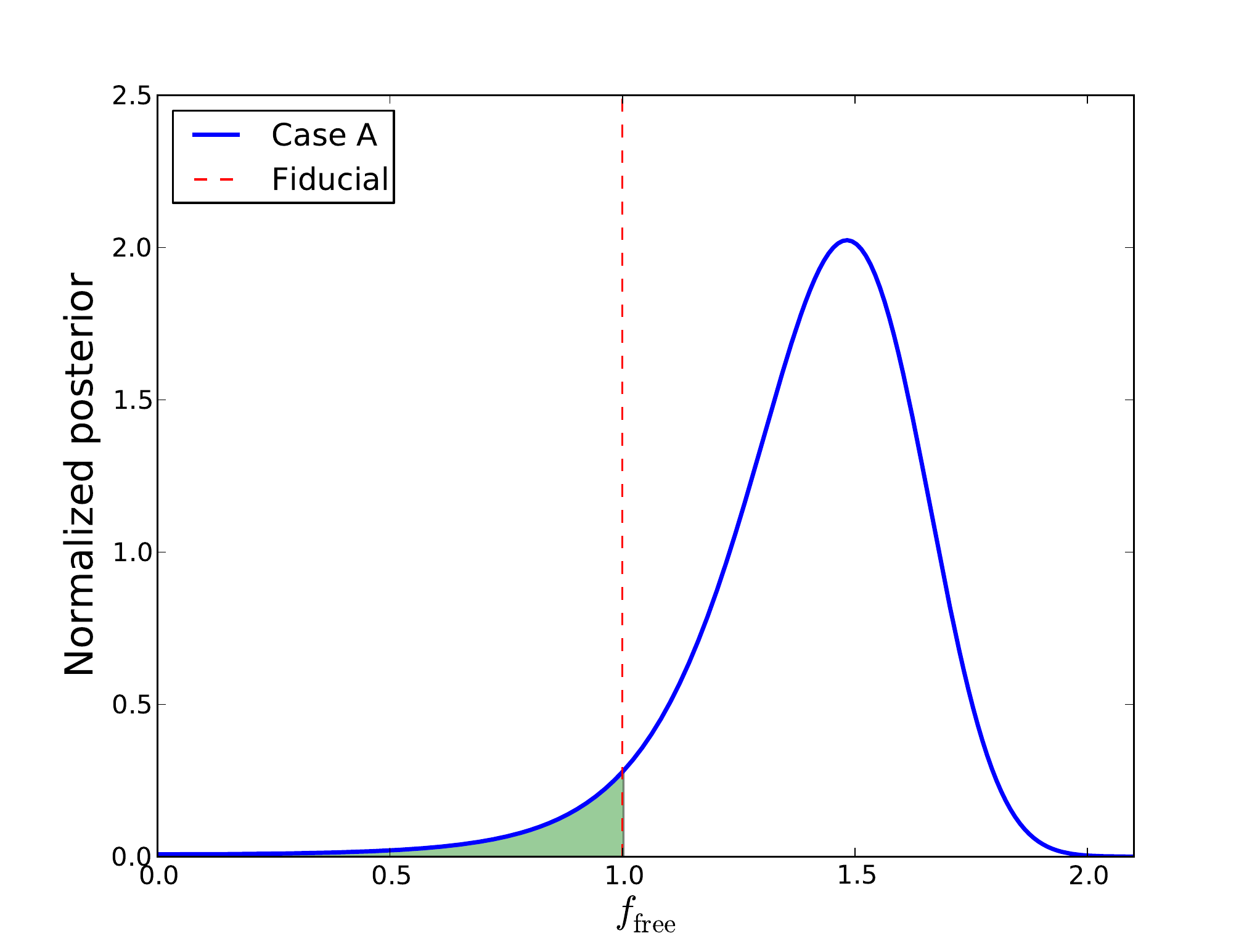}
\caption{Posterior probability for $f_{\rm free}$, obtained by replacing $\mathcal{A}_{\rm kSZ^2}$ with $f_{\rm free}^2$ in Equation \ref{eq.thtemplate} and marginalizing over $b_g$ with no external prior (i.e., Case A from Table~\ref{tab:fits}). The maximum-posterior value is $f_{\rm free}$ = 1.48, slightly larger than the fiducial $f_{\rm free}$ = 1, but we find that the probability of $f_{\rm free} \leq 1$ (the area shaded in green) is 5.4\%. If the posterior were Gaussian, this would correspond to a $1.6 \sigma$ upward fluctuation.}
\label{fig.post_ffree}
\end{figure}

The best-fit kSZ${}^2$ amplitude and galaxy bias are presented in Table \ref{tab:results}.  The results indicate that marginalization over the lensing contribution leads to a degradation of $\approx 15$\% in the error bar on $\mathcal{A}_{\rm kSZ^2}$. The corresponding posterior for $f_{\rm free}$ for our fiducial case (where both the kSZ${}^2$ amplitude and galaxy bias are obtained without using external CMB lensing data) is shown in Figure \ref{fig.post_ffree}. Since $f_{\rm free} \sim \sqrt{\mathcal{A}_{\rm kSZ^2}}$, the posterior is fairly non-Gaussian and shows a considerable negative skewness.  For this reason, our best fit measurement $f_{\rm free} = 1.48$ is only in mild tension with the fiducial value of $f_{\rm free} = 1$, and from the posterior we estimate that the probability of $f_{\rm free} \leq 1$ is 5.4\%, so that if the posterior were Gaussian, this would correspond to a 1.6$\sigma$ upward fluctuation.

\begin{table}[!h]
\begin{tabular}{|l|c|c|}
\hline
Case & $\mathcal{A}_{\rm kSZ^2}$ & $b_g$ \\
\hline \hline
(A): $C_\ell^{T_{\rm clean}^2 \times \delta_g}$ only & $2.18 \pm 0.57$ & $1.10 \pm 0.11$ \\ \hline
(B): $C_\ell^{T_{\rm clean}^2 \times \delta_g}$ and $C_\ell^{\kappa_{\rm CMB} \delta_g}$ & $2.24 \pm 0.50$ & $1.13 \pm 0.02$ \\ \hline
(C): $+10$\% error on $C_\ell^{\kappa_{\rm CMB} \delta_g}$ & $2.21 \pm 0.53$ & $1.11 \pm 0.08$ \\ \hline

\end{tabular}
\caption[]{\label{tab:fits} Best-fit parameters for the kSZ${}^2$--\emph{WISE} galaxies cross-correlation from H16. We include three analysis scenarios: (A) using only the $C_\ell^{T_{\rm clean}^2 \times \delta_g}$ data and marginalizing over the lensing contribution amplitude (i.e., the galaxy bias); (B) including an external prior on the \emph{WISE} galaxy bias from our measurement of $C_\ell^{\kappa_{\rm CMB} \delta_g}$; (C) same as (B), but including an additional $10$\% theoretical systematic error on the $b_g$ constraint from $C_\ell^{\kappa_{\rm CMB} \delta_g}$, due to uncertainties from nonlinear evolution and baryonic physics.}
\label{tab:results}
\end{table}

\section{Challenges}
\label{sec:challenges}
\subsection{Foregrounds}
As in most cross-correlation analyses, there are a number of possible contaminants that have to be carefully scrutinized.  In particular, any emission or imprint of the tracer galaxies\footnote{or emission from other objects that are correlated with the tracer population.} that leaks into the CMB maps will contribute to $C_\ell^{\rm{kSZ}^2 \times \delta_g} $ and could be mistaken for the kSZ$^2$ signal. Since the kSZ signal arises from a Doppler shift in photon energy, it preserves the blackbody spectrum of the CMB, simply producing a small shift in the effective temperature. On the contrary, most other foregrounds give rise to emission that differs considerably from a blackbody at $T_{\rm CMB}$ and can therefore (at least in principle) be separated using multi-frequency analysis (see for example \cite{2015arXiv150205956P}). In particular, while most of the kSZ estimators that require spectroscopic redshifts can be applied to single-frequency CMB temperature maps, our method explicitly requires multi-frequency analysis for foreground subtraction. 

To ensure that the foreground separation is effective a number of null tests can be performed; here we comment on some, but this is far from an exhaustive list.  We have noted before that because of the $\mathbf{v}_e \rightarrow - \mathbf{v}_e$ symmetry, the kSZ contribution to the cross-correlation between tracers and the CMB temperature vanishes, i.e., $\langle \Theta^{\rm kSZ}_f \ \delta_g \rangle = 0$. Checking that this quantity is consistent with zero is therefore a powerful test for the absence of contamination by foregrounds, such as Galactic or extragalactic dust, tSZ, or radio emission.\footnote{There is a small contribution to this correlation due to the ISW effect, which is detectable on large scales and is discussed later.}

Another useful test for dust or radio contamination is to replace one power of the cleaned temperature map in the standard analysis with a tracer of foregrounds (for example the 545 GHz \emph{Planck} map is an excellent tracer of dust emission, while the 30 GHz map traces radio emission). Schematically, we can look at $\langle \Theta^{\rm clean}_f \Theta^{\rm foreground}_f \delta_g \rangle$. While this cross-correlation also contains the kSZ$^2$ signal (in principle), any potential contamination will be greatly enhanced over its contribution to $\langle (\Theta^{\rm clean}_f)^2 \delta_g \rangle$.

The two null tests just described ensure that foregrounds are subtracted correctly \emph{on average}. Spatially varying source properties (such as fluctuations in the spectral index) can lead to a situation where a subset of the sources have been oversubtracted and the others have been undersubtracted; the previous null tests, being linear in $\Theta^{\rm clean}$ are not guaranteed to be sensitive to such a contamination\footnote{they can be sensitive to it if the specific intensity of emission of the sources is correlated with spectral index or other properties.}, but our estimator is, since it is quadratic in $\Theta^{\rm clean}$. One way to test for the latter scenario is to generate mock catalogs in which galaxies are associated with spatially varying emission (with the relevant parameters drawn from a random distribution with scatter matching known source properties). These mock catalogs can then be subjected to the foreground separation pipeline and used in place of $\Theta^{\rm clean}$ in the kSZ$^2$ cross-correlation to estimate the expected amplitude of the effect. All of these null tests were performed in H16.

Note that our method only requires the removal of foregrounds that are correlated with the large-scale structure tracers under consideration.  For example, it is well-known that at high $\ell$, the cosmic infrared background (CIB) is a major contributor to the measured CMB power spectrum, but the bulk emission of the CIB originates from unresolved galaxies at $z \sim 1-3$~\cite{Hill-Spergel2014,2013MNRAS.436.1896A}.

We have shown in H16 that component separation can be used to detect the kSZ$^2$ signal with $S/N \approx 4$ on angular scales up to $\ell \approx 3000$.  We have estimated that the residual contamination is a small fraction of the current statistical uncertainty. It is not yet known how well multi-frequency cleaning techniques will perform at higher $\ell$ and with lower noise levels. This could potentially be the limiting factor in the future performance of this method, and will be the subject of future analysis.

\subsection{Gravitational secondary anisotropies}
There are other secondary CMB anisotropies that preserve the blackbody spectrum of the CMB and therefore cannot be removed by multi-frequency component separation: the contribution from weak lensing and the ISW effect \cite{1967ApJ...147...73S}, as well as its non-linear generalization known as the Rees-Sciama effect \cite{1968Natur.217..511R}. As we have noted in Section \ref{sec:lensing}, the weak lensing contribution can be large and must be accounted for, but its characteristic $\ell$ dependence and the possibility of using external priors allow it to be cleanly separated from the kSZ$^2$ signal. 

Regarding ISW, we should distinguish between the linear and non-linear contributions. The linear part is due to the decay of the gravitational potential on large scales because of the late-time cosmic acceleration. This is a very large-scale effect and detectable at $\ell < 100$ (for a measurement of ISW with \emph{WISE} galaxies, see \cite{Ferraroetal2014, 2016arXiv160403939S}).  For this reason any analysis of the kSZ$^2$ signal should explicitly filter out scales with $\ell$ less than a few hundred. The non-linear contribution is expected to be subdominant to kSZ on all scales with $\ell >$ few hundred.  Perturbation theory and halo model calculations indicate that it is at least two orders of magnitude smaller than kSZ on the scales of interest \cite{2013MNRAS.431.2433M, 2009PhRvD..80f3528S, 2002PhRvD..65h3518C}. If non-perturbative effects are large or the kSZ $S/N$ is large enough (e.g., $\gtrsim$ 100), then this contribution will need to be modeled and accounted for.

\subsection{Theoretical uncertainties}

Finally, we note that the approximations presented here, while more than adequate for the analysis in H16, may need to be improved for the high $S/N$ regime. In particular, we have used fitting functions for the non-linear matter power spectrum and bispectrum, which have a calibration uncertainty $\approx$ 5-10\% \cite{2012ApJ...761..152T, GilMarin:2011ik}, consistent with the level of agreement found when comparing to simulations in Section~\ref{sec:sims}.  In addition, the $C_\ell^{\rm{kSZ}^2 \times \delta_g}$ signal depends steeply on the cosmological parameters, for example scaling as $\sigma_8^{6-7}$~\cite{Doreetal2004}.  Moreover, for the purpose of this work we have assumed that the baryons follow the dark matter exactly on the scales of interest. This should be a good approximation on the scales probed by H16, but it is known not to be the case on small scales.  However, the baryon profile in the outskirts of galaxies and clusters is still very uncertain.  In fact, the small-scale shape of $C_\ell^{\rm{kSZ}^2 \times \delta_g}$ can be used as a probe of the free electron profile, which is sensitive to the effects of feedback and energy injection into the intracluster and intergalactic media (for a measurement of the baryon profile with kSZ and comparison to dark matter, see \cite{2015arXiv151006442S}).

For this analysis, we have used a scale- and redshift-independent galaxy bias, and moreover we have assumed that the shape of the lensing contribution is known exactly, up to a multiplicative constant (that is, the galaxy bias).  While marginalizing over the galaxy bias can mitigate some of the theoretical uncertainties on the amplitude of the lensing term, scale-dependent bias or baryonic effects can introduce systematic effects in the high $S/N$ regime, which may require appropriate treatment in the future.

\subsection{Future directions}
As discussed in the previous section, unmodelled scale-dependent effects in the lensing contribution can potentially mimic the kSZ$^2$ signal and bias the results in the high $S/N$ regime. It is possible to write down estimators that use temperature and/or polarization that are insensitive to the lensing signal, regardless of its amplitude and shape. In particular, CMB polarization is lensed by the same gravitational potential as the CMB temperature, while it receives a negligible contribution from the kSZ effect. Therefore an appropriate combination of temperature and polarization can cancel the lensing signal, while preserving the correct kSZ$^2$ amplitude.

Another improvement that can be implemented in future analyses is optimal redshift weighting of the projected tracer field in Equation \ref{eq:delta_gproj}, which has not been considered in this work. Ref.~\cite{Doreetal2004} shows that the peak differential contribution to the kSZ$^2$ signal comes from $z \sim 0.5$, and the \emph{WISE} galaxy distribution is fairly well matched to the signal redshift distribution.  Optimal weighting should especially benefit surveys for which the source distribution is peaked at higher redshift or with very extended tails. For example, the \emph{SPHEREx} experiment might benefit from downweighting the high-redshift population tail and it may be possible to obtain higher statistical significance than that predicted in Table \ref{tab:forecast}.

These points will be explored in future work.

\section{Conclusions}
\label{sec:conclusions}
We have revisited a kSZ estimator based on projected fields, which does not require expensive spectroscopic data. This will allow the use of large, full-sky imaging catalogs for kSZ measurements, yielding accurate determinations of the low-redshift baryon abundance and the free electron distribution associated with galaxies and clusters.  In a companion paper (H16), we have shown that this method is already competitive with other kSZ approaches when applied to current data, allowing a detection of the kSZ signal with $S/N \approx 4$ by combining \emph{Planck} and \emph{WMAP} microwave temperature maps with a \emph{WISE} galaxy catalog. If foreground cleaning methods in future experiments are effective at separating the CMB blackbody component from other microwave sky signals, we forecast kSZ measurements with $S/N > 100$ for \emph{Advanced ACTPol} and \emph{CMB-S4}. This will allow precision measurements of both the abundance and profile of the baryons associated with the tracer sample.  Since both of these properties are expected to vary with mass and redshift, the tracer population can be split into multiple samples that can be compared to high precision. In addition, other properties such as color, star formation rate, or AGN activity are expected to influence the gas distribution, and comparing the kSZ$^2$ signal from multiple different tracer populations will shed light on galaxy evolution and feedback processes.  When combined with tSZ measurements of the same objects, the gas temperature, density, and pressure of the intergalactic medium can be simultaneously inferred, providing information about the amount of energy injection.

It is also important to point out that while for concreteness we have shown forecasts for ``galaxy overdensity'' as our tracer, any tracer of the late-time density can be used in this approach. In particular we expect interesting measurements when using galaxy lensing as a tracer (our measurement will then probe the matter-gas correlation), or 21 cm observations (to probe the ionized-neutral gas correlation).

Finally, these measurements will soon complement kSZ measurements obtained from the small-scale CMB power spectrum \cite{2012ApJ...756...65Z, 2014JCAP...08..010C}, and will be useful to disentangle the contributions due to late-time structure from those produced during ``patchy'' cosmic reionization.

\acknowledgments
We are grateful to Olivier Dor{\'e}, Zoltan Haiman, Emmanuel Schaan, Blake Sherwin and Kendrick Smith for very useful conversations.  We also thank the LGMCA team for publicly releasing their CMB maps.  SF, JCH and DNS acknowledge support from NASA Theory Grant NNX12AG72G and NSF AST-1311756. SF thanks the Miller Institute for Basic Research in Science at the University of California, Berkeley for support.  This work was partially supported by a Junior Fellow award from the Simons Foundation to JCH.   NB acknowledges support from the Lyman Spitzer Fellowship.  JL is supported by NSF grant AST-1210877.  Some of the results in this paper have been derived using the HEALPix package~\cite{Gorskietal2005}. This publication makes use of data products from the \emph{Wide-field Infrared Survey Explorer}, which is a joint project of the University of California, Los Angeles, and the Jet Propulsion Laboratory/California Institute of Technology, funded by the National Aeronautics and Space Administration.

\begin{appendix}
\section{Assumptions about \emph{WISE} and \emph{SPHEREx}}
\begin{figure}[ht]
\centering
\includegraphics[width=0.5\textwidth]{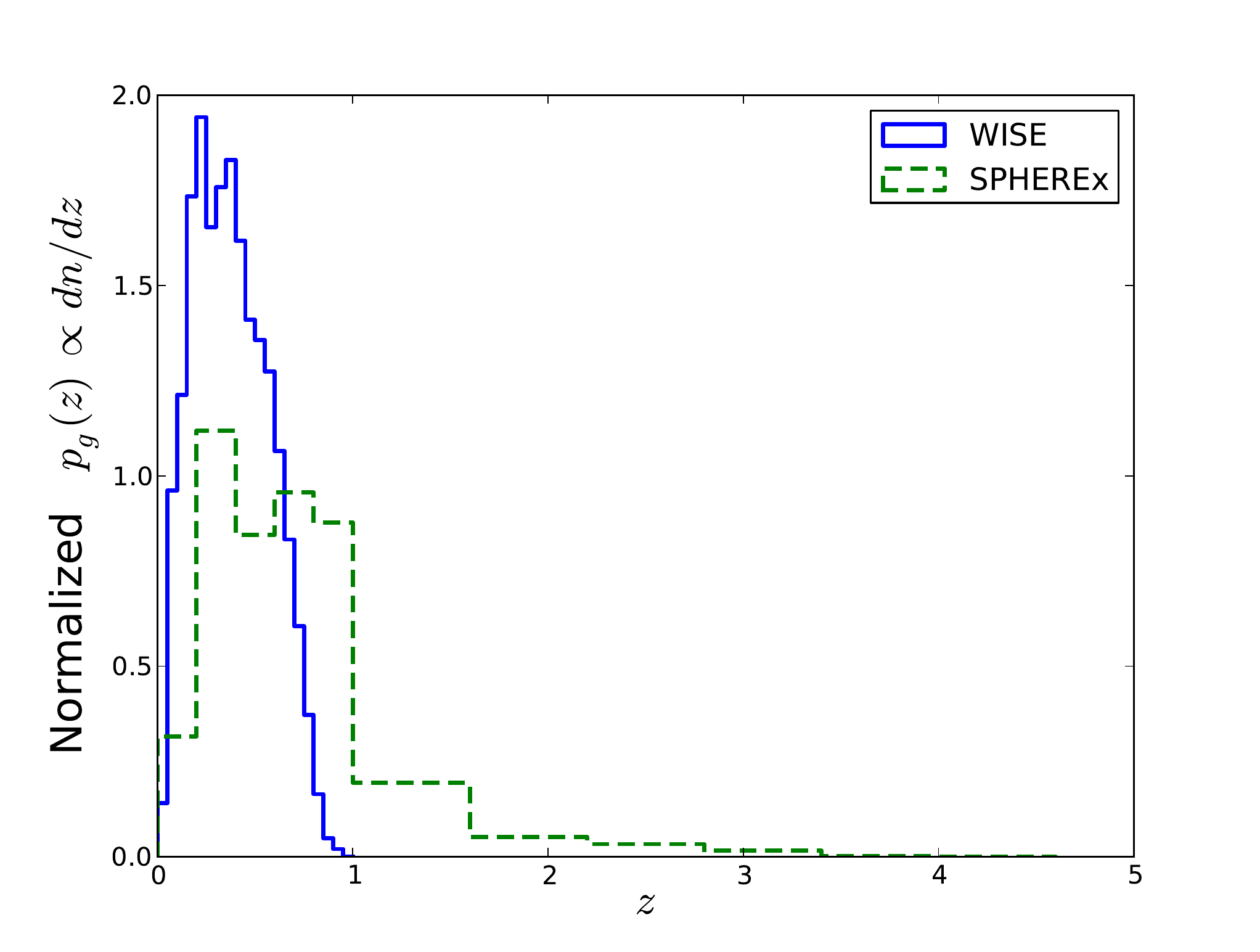}
\caption{Normalized redshift distributions for \emph{WISE} and \emph{SPHEREx} galaxies. The weight $p_g(z)$ is related to $p_g(\eta)$ in Equation \ref{eq.Wg} by $p_g(\eta) = H(z) p_g(z)$.}
\label{fig.dndz}
\end{figure}

In this appendix we show the assumed redshift distributions for \emph{WISE} and \emph{SPHEREx} galaxies, derived from Refs.~\cite{2013AJ....145...55Y} and \cite{2014arXiv1412.4872D}, respectively (see Figure~\ref{fig.dndz}).  For \emph{WISE} we have approximately 50 million galaxies over half of the sky, while on the same footprint, the full \emph{SPHEREx} galaxy catalog is predicted to have about 290 million objects.

The galaxy bias is assumed constant for \emph{WISE}, while for \emph{SPHEREx} we use the (redshift-dependent) bias model from \cite{2014arXiv1412.4872D}.
\end{appendix}

\newpage

\begin{thebibliography}{3}

\bibitem[Hinshaw et al.(2013)]{Hinshawetal2013} Hinshaw, G., Larson, D., Komatsu, E., et al.\ 2013, \apjs, 208, 19 
\bibitem[Planck Collaboration et al.(2015)]{2015arXiv150201582P} Planck Collaboration, Adam, R., Ade, P.~A.~R., et al.\ 2015, arXiv:1502.01582 
\bibitem[Steigman(2007)]{Steigman2007} Steigman, G.\ 2007, Annual Review of Nuclear and Particle Science, 57, 463 
\bibitem[Fukugita et al.(1998)]{Fukugitaetal1998} Fukugita, M., Hogan, C.~J., \& Peebles, P.~J.~E.\ 1998, \apj, 503, 518 
\bibitem[Bregman(2007)]{Bregman2007} Bregman, J.~N.\ 2007, \araa, 45, 221 
\bibitem[Cen \& Ostriker(2006)]{2006ApJ...650..560C} Cen, R., \& Ostriker, J.~P.\ 2006, \apj, 650, 560 
\bibitem[Werk et al.(2014)]{2014ApJ...792....8W} Werk, J.~K., Prochaska, J.~X., Tumlinson, J., et al.\ 2014, \apj, 792, 8 
\bibitem[Bonamente et al.(2016)]{2016MNRAS.tmp...78B} Bonamente, M., Nevalainen, J., Tilton, E., et al.\ 2016, \mnras
\bibitem[Sunyaev \& Zeldovich(1972)]{Sunyaev-Zeldovich1972} Sunyaev, R.~A., \& Zel'dovich, Y.~B.\ 1972, Comments Astrophys.~ Space Phys., 4, 173 
\bibitem[Sunyaev \& Zeldovich(1980)]{Sunyaev-Zeldovich1980} Sunyaev, R.~A., \& Zeldovich, I.~B.\ 1980, \araa, 18, 537 
\bibitem[Ostriker \& Vishniac(1986)]{Ostriker-Vishniac1986} Ostriker, J.~P., \& Vishniac, E.~T.\ 1986, \apjl, 306, L51 
\bibitem[Bhattacharya \& Kosowsky(2008)]{2008PhRvD..77h3004B} Bhattacharya, S., \& Kosowsky, A.\ 2008, \prd, 77, 083004 
\bibitem[Hand et al.(2012)]{Handetal2012} Hand, N., Addison, G.~E., Aubourg, E., et al.\ 2012, Physical Review Letters, 109, 041101 
\bibitem[Planck Collaboration et al.(2015)]{2015arXiv150403339P} Planck Collaboration, Ade, P.~A.~R., Aghanim, N., et al.\ 2015, arXiv:1504.03339 
\bibitem[Hern{\'a}ndez-Monteagudo et al.(2015)]{2015PhRvL.115s1301H} Hern{\'a}ndez-Monteagudo, C., Ma, Y.-Z., Kitaura, F.~S., et al.\ 2015, Physical Review Letters, 115, 191301 
\bibitem[Schaan et al.(2015)]{2015arXiv151006442S} Schaan, E., Ferraro, S., Vargas-Maga{\~n}a, M., et al.\ 2015, arXiv:1510.06442 
\bibitem[Soergel et al.(2016)]{2016arXiv160303904S} Soergel, B., Flender, S., Story, K.~T., et al.\ 2016, arXiv:1603.03904 
\bibitem[Mroczkowski et al.(2012)]{Mroczkowskietal2012} Mroczkowski, T., Dicker, S., Sayers, J., et al.\ 2012, \apj, 761, 47 
\bibitem[Sayers et al.(2013)]{Sayersetal2013} Sayers, J., Mroczkowski, T., Zemcov, M., et al.\ 2013, \apj, 778, 52 
\bibitem[Ho et al.(2009)]{2009arXiv0903.2845H} Ho, S., Dedeo, S., \& Spergel, D.\ 2009, arXiv:0903.2845
\bibitem[Shao et al.(2011)]{2011MNRAS.413..628S} Shao, J., Zhang, P., Lin, W., Jing, Y., \& Pan, J.\ 2011, \mnras, 413, 628
\bibitem[Li et al.(2014)]{2014MNRAS.443.2311L} Li, M., Angulo, R.~E., White, S.~D.~M., \& Jasche, J.\ 2014, \mnras, 443, 2311 
\bibitem[Ferreira et al.(1999)]{1999ApJ...515L...1F} Ferreira, P.~G., Juszkiewicz, R., Feldman, H.~A., Davis, M., \& Jaffe, A.~H.\ 1999, \apjl, 515, L1 
\bibitem[Keisler \& Schmidt(2013)]{2013ApJ...765L..32K} Keisler, R., \& Schmidt, F.\ 2013, \apjl, 765, L32 
\bibitem[Flender et al.(2015)]{2015arXiv151102843F} Flender, S., Bleem, L., Finkel, H., et al.\ 2015, arXiv:1511.02843 
\bibitem[Wright et al.(2010)]{2010AJ....140.1868W} Wright, E.~L., Eisenhardt, P.~R.~M., Mainzer, A.~K., et al.\ 2010, \aj, 140, 1868-1881 
\bibitem[Dor{\'e} et al.(2004)]{Doreetal2004} Dor{\'e}, O., Hennawi, J.~F., \& Spergel, D.~N.\ 2004, \apj, 606, 46 
\bibitem[DeDeo et al.(2005)]{DeDeoetal2005} DeDeo, S., Spergel, D.~N., \& Trac, H.\ 2005, arXiv:astro-ph/0511060 
\bibitem[Hill et al.(2016)]{2016arXiv160301608H} Hill, J.~C., Ferraro, S., Battaglia, N., Liu, J., \& Spergel, D.~N.\ 2016, arXiv:1603.01608 
\bibitem[Bobin et al.(2015)]{Bobinetal2015} Bobin, J., Sureau, F., \& Starck, J.\ 2015, arXiv:1511.08690 
\bibitem[Bennett et al.(2013)]{Bennettetal2013} Bennett, C.~L., Larson, D., Weiland, J.~L., et al.\ 2013, \apjs, 208, 20 
\bibitem[Planck Collaboration et al.(2015)]{Planck2015params} Planck Collaboration, Ade, P.~A.~R., Aghanim, N., et al.\ 2015, arXiv:1502.01589  
\bibitem[Battaglia et al.(2010)]{BBPSS2010} Battaglia, N., Bond, J.~R., Pfrommer, C., Sievers, J.~L., \& Sijacki, D.\ 2010, \apj, 725, 91
\bibitem[Limber(1953)]{1953ApJ...117..134L} Limber, D.~N.\ 1953, \apj, 117, 134 
\bibitem{Hahn:2014lca} Hahn, O., Angulo, R.~E., \& Abel, T.\ 2015, \mnras, 454, 3920 
\bibitem{GilMarin:2011ik} Gil-Mar{\'{\i}}n, H., Wagner, C., Fragkoudi, F., Jimenez, R., \& Verde, L.\ 2012, \jcap, 2, 047 
\bibitem[Dor{\'e} et al.(2014)]{2014arXiv1412.4872D} Dor{\'e}, O., Bock, J., Ashby, M., et al.\ 2014, arXiv:1412.4872   
\bibitem[Henderson et al.(2015)]{Henderson2015} Henderson, S.~W., Allison, R., Austermann, J., et al.\ 2015, arXiv:1510.02809 
\bibitem[Lewis \& Challinor(2006)]{2006PhR...429....1L} Lewis, A., \& Challinor, A.\ 2006, \physrep, 429, 1 
\bibitem[Sehgal et al.(2010)]{2010ApJ...709..920S} Sehgal, N., Bode, P., Das, S., et al.\ 2010, \apj, 709, 920 
\bibitem[Springel(2005)]{2005MNRAS.364.1105S} Springel, V.\ 2005, \mnras, 364, 1105 
\bibitem[Pfrommer et al.(2006)]{2006MNRAS.367..113P} Pfrommer, C., Springel, V., En{\ss}lin, T.~A., \& Jubelgas, M.\ 2006, \mnras, 367, 113 
\bibitem[En{\ss}lin et al.(2007)]{2007A&A...473...41E} En{\ss}lin, T.~A., Pfrommer, C., Springel, V., \& Jubelgas, M.\ 2007, \aap, 473, 41 
\bibitem[Jubelgas et al.(2008)]{2008A&A...481...33J} Jubelgas, M., Springel, V., En{\ss}lin, T., \& Pfrommer, C.\ 2008, \aap, 481, 33 
\bibitem[Springel \& Hernquist(2003)]{2003MNRAS.339..289S} Springel, V., \& Hernquist, L.\ 2003, \mnras, 339, 289 
\bibitem[Battaglia et al.(2012)]{BBPS2012b} Battaglia, N., Bond, J.~R., Pfrommer, C., \& Sievers, J.~L.\ 2012, \apj, 758, 75 
\bibitem[Heymans et al.(2012)]{Heymansetal2012} Heymans, C., Van Waerbeke, L., Miller, L., et al.\ 2012, \mnras, 427, 146 
\bibitem[Battaglia et al.(2015)]{BHM2014} Battaglia, N., Hill, J.~C., \& Murray, N.\ 2015, \apj, 812, 154 
\bibitem[Park et al.(2013)]{2013ApJ...769...93P} Park, H., Shapiro, P.~R., Komatsu, E., et al.\ 2013, \apj, 769, 93 
\bibitem[Takahashi et al.(2012)]{2012ApJ...761..152T} Takahashi, R., Sato, M., Nishimichi, T., Taruya, A., \& Oguri, M.\ 2012, \apj, 761, 152 
\bibitem[Bobin et al.(2014)]{Bobinetal2014} Bobin, J., Sureau, F., Starck, J.-L., Rassat, A., \& Paykari, P.\ 2014, \aap, 563, A105 
\bibitem[Bobin et al.(2013)]{Bobinetal2013} Bobin, J., Starck, J.-L., Sureau, F., \& Basak, S.\ 2013, \aap, 550, A73 
\bibitem[Yan et al.(2013)]{2013AJ....145...55Y} Yan, L., Donoso, E., Tsai, C.-W., et al.\ 2013, \aj, 145, 55 
\bibitem[Ferraro et al.(2014)]{Ferraroetal2014} Ferraro, S., Sherwin, B.~D., \& Spergel, D.~N.\ 2014, arXiv:1401.1193 
\bibitem[Jarrett et al.(2011)]{2011ApJ...735..112J} Jarrett, T.~H., Cohen, M., Masci, F., et al.\ 2011, \apj, 735, 112 
\bibitem[Planck Collaboration et al.(2014)]{Planck2013lensing} Planck Collaboration, Ade, P.~A.~R., Aghanim, N., et al.\ 2014, \aap, 571, A17 
\bibitem[Planck Collaboration et al.(2015)]{Planck2015lensing} Planck Collaboration, Ade, P.~A.~R., Aghanim, N., et al.\ 2015, arXiv:1502.01591 
\bibitem[Planck Collaboration et al.(2015)]{2015arXiv150205956P} Planck Collaboration, Adam, R., Ade, P.~A.~R., et al.\ 2015, arXiv:1502.05956 
\bibitem[Hill \& Spergel(2014)]{Hill-Spergel2014} Hill, J.~C., \& Spergel, D.~N.\ 2014, \jcap, 2, 30
\bibitem[Addison et al.(2013)]{2013MNRAS.436.1896A} Addison, G.~E., Dunkley, J., \& Bond, J.~R.\ 2013, \mnras, 436, 1896 
\bibitem[Sachs \& Wolfe(1967)]{1967ApJ...147...73S} Sachs, R.~K., \& Wolfe, A.~M.\ 1967, \apj, 147, 73 
\bibitem[Rees \& Sciama(1968)]{1968Natur.217..511R} Rees, M.~J., \& Sciama, D.~W.\ 1968, \nat, 217, 511 
\bibitem[Shajib \& Wright(2016)]{2016arXiv160403939S} Shajib, A.~J., \& Wright, E.~L.\ 2016, arXiv:1604.03939 
\bibitem[Merkel \& Sch{\"a}fer(2013)]{2013MNRAS.431.2433M} Merkel, P.~M., \& Sch{\"a}fer, B.~M.\ 2013, \mnras, 431, 2433 
\bibitem[Smith et al.(2009)]{2009PhRvD..80f3528S} Smith, R.~E., Hern{\'a}ndez-Monteagudo, C., \& Seljak, U.\ 2009, \prd, 80, 063528 
\bibitem[Cooray(2002)]{2002PhRvD..65h3518C} Cooray, A.\ 2002, \prd, 65, 083518 
\bibitem[Zahn et al.(2012)]{2012ApJ...756...65Z} Zahn, O., Reichardt, C.~L., Shaw, L., et al.\ 2012, \apj, 756, 65 
\bibitem[Calabrese et al.(2014)]{2014JCAP...08..010C} Calabrese, E., Hlozek, R., Battaglia, N., et al.\ 2014, \jcap, 8, 010 
\bibitem[G{\'o}rski et al.(2005)]{Gorskietal2005} G{\'o}rski, K.~M., Hivon, E., Banday, A.~J., et al.\ 2005, \apj, 622, 759 

\end{thebibliography}
\end{document}